\documentclass{article}
\usepackage[utf8]{inputenc}
\usepackage{lmodern}
\usepackage[T1]{fontenc}
\usepackage{lscape}%A mettre en plus
\usepackage{amsmath}	% les packages asm = american society of mathematics => symboles etc.
\usepackage{amssymb}
\usepackage{amsthm}
\usepackage{cases}

 \usepackage[demo]{graphicx}
 
\usepackage{caption}

\usepackage{multirow}
\usepackage{amsfonts}
\usepackage{float} 
\usepackage{diagbox}
\usepackage{subfigure}%! LaTeX Error: Command \c@subfigure already defined.Or name \end... illegal, see p.192 of the manual.See the LaTeX manual or LaTeX Companion for explanation.Type H <return> for immediate help.... \newcounter{subfigure}
\usepackage{fullpage}
\usepackage{url}
\usepackage{rotating}
\usepackage{eurosym}%symbole?
\usepackage{wrapfig}%image à droite
\usepackage[final]{pdfpages} %inclusion de pdf
\usepackage{epstopdf} %pour les fichiers eps
\usepackage[a4paper]{geometry}
\usepackage[nottoc]{tocbibind} %Adds "References" to the table of contents
\usepackage{placeins}
\usepackage{float}

\usepackage{listings}
\usepackage{color, colortbl}

\usepackage{epstopdf}

\usepackage {tikz}
\usetikzlibrary {positioning}

\definecolor{mygreen}{rgb}{0,0.6,0}
\definecolor{mygray}{rgb}{0.5,0.5,0.5}
\definecolor{mymauve}{rgb}{0.58,0,0.82}
\definecolor{airforceblue}{rgb}{0.36, 0.54, 0.66}
\definecolor{celadon}{rgb}{0.67, 0.88, 0.69}
\definecolor{coralpink}{rgb}{0.97, 0.51, 0.47}
\definecolor{chromeyellow}{rgb}{1.0, 0.65, 0.0}
\definecolor{pastelpurple}{rgb}{0.7, 0.62, 0.71}
\definecolor{timberwolf}{rgb}{0.86, 0.84, 0.82}
\definecolor{sandstorm}{rgb}{0.93, 0.84, 0.25}
\definecolor {processblue}{cmyk}{0.96,0,0,0}

\title{\textbf{Role model detection using \\ low rank similarity matrix}}
\author{
CHENG Sibo,\, LAURENT Adissa\footnote{Catholic University of Louvain, Departement of Mathematical Engineering, Avenue Georges Lemaitre 4, B-1348 Louvain-la-Neuve,Belgium. This article is written in the case of a project  guided by Professor P.VAN DOOREN(third author),S.CHENG and A.LAURENT contributed equally to the work.},\,\,VAN DOOREN Paul\footnote{The work of third author is partly supported by the Belgian Network DYSCO (Dynamical Systems,Control, and Optimization),funded by the Inter-university Attraction Poles Programme,initialed by the Belgian State,Science Policy Office.The scientific responsibility restes with its authors} 
}
\date{ Avril 2016}
\providecommand{\keywords}[1]{\textbf{\textit{Key words---}} #1}

\begin{document}

\maketitle

%\tableofcontents

%\newpage

\begin{abstract}
    Computing meaningful clusters of nodes is crucial
to analyse large networks. In this paper, we apply new clustering methods to improve the computational time. We use the properties of the adjacency matrix to obtain better role extraction. We also define a new non-recursive similarity measure and compare its results with the ones obtained with Browet's similarity measure.  We will show the extraction of the different roles with a linear time complexity. Finally, we test our algorithm with real data structures and analyse the limit of our algorithm.
\end{abstract}

\section{Introduction}
Over the years, the development of storage capacities has allowed to collect huge amounts of data (food-webs, human interactions, word classification...). This data can be represented as network structures with agents stored as nodes and their mutual information as edges. Graph theory then allows us to understand and analyse those large networks and retrieve the graph structure. An important structure in a graph of communication is its different roles. A role is a group of nodes sharing similar behaviour or flow patterns within the network. To derive the role model we will use pairwise similarity measures based on the similarity matrix $S$. The measure we will use principally is a low rank iterative scheme proposed by Browet in \cite{IEEEhowto:Browet1} and \cite{IEEhowto:BrowetThesis}. It computes the matrix $X$ of a low rank factorisation $S = X X^T$. We will then introduce a non-iterative and faster measure based on the one proposed by Browet. When this is done, one needs to apply a community detection algorithm on the matrix $S$ to find the partition into the different subsets and detect the roles afterwards. However, $S$ is dense and contains $\mathcal{O}(n^2)$ non-zero elements. To efficiently implement the community detection, which is crucial on large databases, we will implicitly work on the lowest rank factor $X$ of size $\mathcal{O}(n)$ elements. For that purpose, we will use K-means algorithm on $X$ in the case the number of clusters k is given. To improve the classification, we will use properties of the adjacency matrix to decide whether restarting K-means algorithm is necessary and if the number of chosen clusters is adequate. If $k$ has not previously been given, we make the assumption to have the knowledge of an upper bound $r$ of $k$. Several methods based on the hierarchical classification and singular value projection will then be implemented in order to find the correct number $k$. Finally, we will illustrate our results on Erdos-Renyi random graphs and on real data structures.

\section{Model and methods}
\subsection{Similarity measures}
\begin{figure}[!h]
\centering
\includegraphics[width=4.5in]{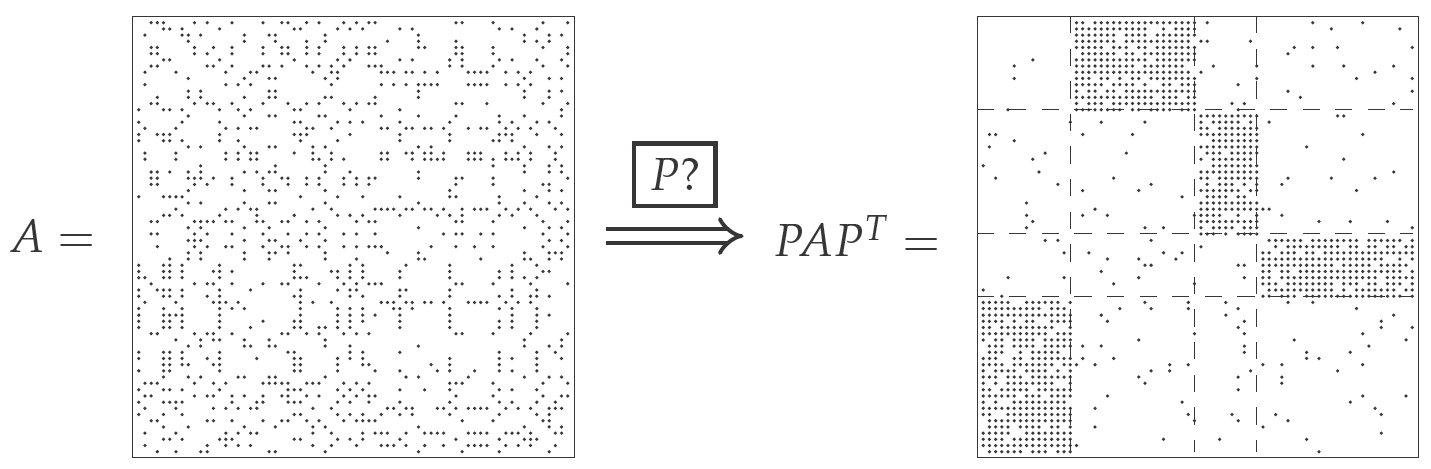}
\caption{Bloc modelling, Adapted from “Algorithms for community and role detection in networks,” by A. Browet and P. Van Dooren, 2013}
\label{fig_perm}
\end{figure}

We consider a directed graph $G_A(V,E)$ with $V$ the set of vertices and $E$ the set of edges. A directed graph is such that an edge $(i,j)$ has a source $i$ and a destination $j$. 
The adjacency matrix $A \in \mathbb{R}^{n \times n}$ of the graph is defined as 
\begin{equation}
    A(i,j) =  
    \left\{
    \begin{split}
    1 \text{ if } (i,j) \in E \\ 
    0 \text{ otherwise }
    \end{split}
  \right.
\end{equation}
The role extraction problem or bloc modelling consist of finding a permutation matrix $P$ such that the edges of the relabelled graph, associated with the adjacency matrix $P A P^T$, are mainly concentrated within blocks as shown in figure \ref{fig_perm}.
The role extraction problem is based on the assumption that nodes can be clustered according to a suitable measure of equivalence. Pairwise self-similarity measure compares each node of the input graph with all the nodes of the same graph. It thus computes the similarity between each pair of nodes and then cluster highly similar nodes together.

\begin{figure}[!h]
\centering
\includegraphics[width=5.3in]{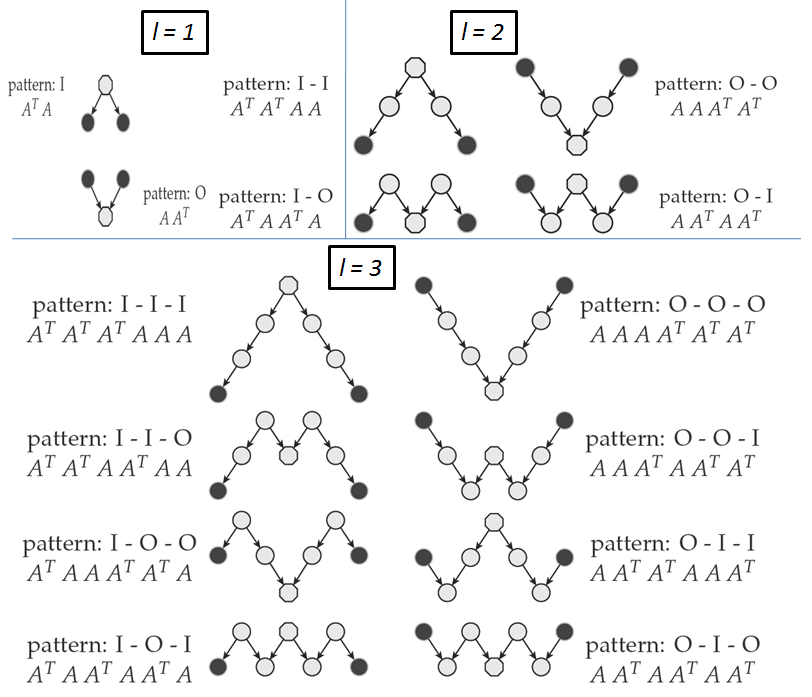}
\caption{All the different neighbourhood patterns, up to length 3, captured by similarity measure $S_{i,j}$ with the source nodes $i$ and $j$ represented as dark circles and the target node represented as light gray circles., Adapted from “Algorithms for community and role detection in networks,” by A. Browet and P. Van Dooren, 2013 }
\label{fig_parentChild}
\end{figure}
If we define a neighbourhood pattern of length $l$ for a node as the sequence of length $l$ of incoming (I) and outgoing (O) edges starting from the source node and reaching the target node (see figure \ref{fig_parentChild}), one similarity criterion is to ask a similar pair of nodes to have the same neighbourhood patterns in common. In other words, a pair of similar nodes should reach many common targets with the same neighbourhood patterns and do so for patterns of various lengths.
To illustrate this concept, let us first define a few terms. In a directed graph, a neighbour $j$ of a node $i$ is called a child when $(i,j) \in E$ and a parent when $(j,i) \in E$. For a given node $i$, one can thus compute the number of parents or in-degree $k_i^{in}$ and the number of children or out-degree $k_i^{out}$. The vectors of out and in-degrees are given by
\begin{align}
    k^{out} = A \textbf{1} && k^{in} = A^T \textbf{1}
\end{align}
where $\mathbf{1}$ is a vector of 1's of appropriate dimension. The number of common children between a pair of nodes $(i,j)$ is thus given by 
\begin{equation}
    [A A^T]_{i,j} = \# \left\{k | i \rightarrow k \text{ and } j \rightarrow k \right\}
\end{equation}
and the number of common parents by
\begin{equation}
    [A^T A]_{i,j} = \# \left\{k | k \rightarrow i \text{ and } k \rightarrow j \right\}
\end{equation}
The total number of common parents and children for nodes $i$ and $j$ are thus given by $[A^T A + A A^T]_{i,j}$. Generalising it for a pattern of length $l$, one obtains that the number of common target nodes for neighbourhood patterns of length $l$ is given by 
\begin{align}
    T_l &= A T_{l-1} A^T + A^T T_{l-1} A \\
    T_1 &= A^T A + A A^T
\end{align}
Browet then defines its pairwise node similarity measure as the weighted sum of the number of common target nodes using neighbourhood patterns of any length
\begin{equation}
    S = \sum_{l=1}^{\infty} \beta^{2(l-1)} T_l = \sum_{l=1}^{\infty} \beta^{2(l-1)} \Gamma_A^l [I]
\end{equation}
where 
\begin{equation}
    \Gamma_A : \mathbb{R}^{n \times n} \rightarrow \mathbb{R}^{n \times n}: \Gamma_A[X] = A X A^T + A X A
\end{equation}
and $\beta \in \mathbb{R}$ is a scaling parameter to balance the relative importance of long neighbourhood
patterns with respect to short neighbourhood patterns since the number of common targets tends to naturally grow when using longer patterns. 
The similarity matrix $S$ can be computed as the fixed point solution of 
\begin{equation} \label{itSchemeEq}
    S_{k+1} = \Gamma_A[I + \beta^2 S_k]
\end{equation}
If we initialise the sequence with $S_0 = 0$, the iteration can be written as
\begin{align}
    S_{k + 1} &= S_1 + \beta^2 \Gamma_A[S_k] \\
    S_1 &= A A^T + A^T A
\end{align}
If $\beta$ is small enough, the sequence converges to
\begin{equation}
    S^{*} = S_1 + \beta^2 (ASA^T + A^T S A)
\end{equation}
using the property of the Kronecker product, the fixed point solution can be written as
\begin{equation}
    vec(S^{*}) = [I - \beta^2 (A \otimes A^T + (A \otimes A)^T]^{-1} vec(S_1)
\end{equation}
To ensure convergence, one can choose $\beta$ small enough.
However, even if $\beta$ is small enough to ensure convergence, it might be impossible to compute the fixed point solution of equation \ref{itSchemeEq} because of the increasing computational cost and memory requirement. Indeed, even if $A$ is sparse, the matrix $S_k$ tends to fill in as $k$ increases and each iteration of \ref{itSchemeEq} is $\mathcal{O}(n^3)$.

Therefore, Browet defines a low rank similarity approximation of rank at most $r$ of $S^{*}$ as 
\begin{equation}\label{itSchemeRedEq}
    S_{k+1}^{(r)} = \Pi^{(r)} [S_1^{(r)} + \beta^2 \Gamma_A [S_k^{(r)}] ] = X_{k+1} X_{k+1}^T
\end{equation}
where $X_k \in \mathbb{R}^{n \times r}$ and $\Pi^{(r)} [.]$ is the best low-rank projector on the dominant subspace of dimension at most $r$ which can be computed using a truncated singular value decomposition (SVD). $S_1^{(r)}$ is the best low-rank approximation of $S_1$ which can be written as
\begin{equation}
    S_1 = [A | A^T][A | A^T]^T
\end{equation}
where $[A|A^T]$ is the horizontal concatenation of $A$ and $A^T$. The singular value decomposition of this concatenation can be computed as
\begin{equation}
    [A | A^T] = U_1 \Sigma_1 V_1^T + U_2 \Sigma_2 V_2^T
\end{equation}
where the columns of the unitary matrix $U_1 \in \mathbb{R}^{n \times r} $ span the dominant
subspace of dimension at most $r$ of $[A|A^T]$ and $\Sigma_1 \in \mathbb{R}^{r \times r}$ is the diagonal matrix of the dominant singular values, $\Sigma_1(i, i) > \Sigma_2(j, j), \forall i, j$. This leads to 
\begin{equation}
    [A|A^T] [A|A^T]^T = U_1 \Sigma_1^2 U_1^T + U_2 \Sigma_2 U_2^T
\end{equation}
which implies the low rank projection of $S_1$ is given by
\begin{equation}
    S_1^{(r)} = U_1 \Sigma_1^2 U_1^T = X_1 X_1^T
\end{equation}
To compute each iterative solution of equation \ref{itSchemeRedEq}, one can see that
\begin{align}
    S_1^{(r)} + \beta^2 \Gamma_A [S_k^{(r)} ] &= Y_k Y_k^T \\
    Y_k &= [X_1 | \beta A X_k | \beta A^T X_k] \\
    X_{k+1} X_{k+1}^T &= \Pi^{(r)}[Y_k Y_k^T]
\end{align}
To efficiently compute $X_{k+1}$, we first apply a QR factorisation to $Y_k = Q_k R_k$, then compute a truncated SVD of rank at most $r$ of $R_k$ such that $R_k = \mathcal{U}_k \Omega_k \mathcal{V}_k$ and finally compute 
\begin{equation}
    X_{k+1} = Q_k \mathcal{U}_k \Omega_k
\end{equation}
Browet proved in his thesis that the low rank iteration converges to
\begin{align}
    S^{(r)} &= X X^T = \Pi^{(r)}[Y Y^T]\\
    Y Y^T   &= [X_1 | \beta A X | \beta A^T X]
\end{align}
if the spectral gap of $YY^T$ at the $r$th eigenvalue is sufficiently large and choosing $\beta$ sufficiently small. One way to ensure convergence is to choose 
\begin{equation}
    \beta^2 < \frac{1}{ || A \otimes A + A^T \otimes A^T ||_F \left( \frac{8 || \Sigma^2 ||}{\Sigma_{k}^2 - \sigma_{1}^2} + 1 \right) }
\end{equation}

\subsubsection{A new method to define the similarity matrix}
We will now introduce a new definition of similarity matrices which we intuitively created but found later in the thesis of Thomas P. Cason. This method is called the Salton index method. Its purpose is to find a new similarity matrix which can show more efficiently that if two nodes belong to one same cluster : it does only take into account "brothers" relationships (two nodes are "brothers" if they possess a large number of children and parent nodes in common). Two "brother" nodes should belong to one same cluster. Thus, the method of Browet seems to consider too much information. It considers further family relatives by checking if two nodes possess same grandparents or grandchildren, etc. Indeed, when two nodes possess the same parents and children, they automatically possess the same grand parents and grand children.

We expect to have values near to 0 or 1 in the similarity matrix by examining only the number of common parents and children they possess. If the common number of children and parents are already known, the information of common grandparents and grandchildren become useless to decide if the two nodes belong to one same cluster and risk to cloud the similarity matrix, making the block identification harder. The common grandparents who do not come from the same parents is considered as a perturbation. 

We start by taking $\beta=0$ in Browet's method,
\begin{equation} \label{simi_Sibo}
    S=AA^{T}+A^{T}A
\end{equation}
For this similarity matrix, we only consider the common number of parents and children but we do not take into account the total number of and children that each node possesses. This may lead to incorrect clusters.
\begin{figure}
    \centering
    \includegraphics[scale=0.5]{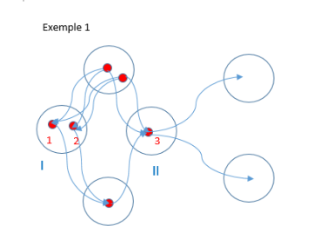}
    \includegraphics[scale=0.6]{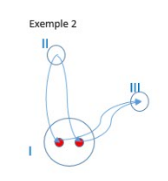}
    \caption{Example of block structure}
    \label{fig:example1}
\end{figure}
For the instance in the figure \ref{fig:example1} Example 1, one can observe that  node 1 and node 2 have the same number of common parents and children as node 1 and node 3. Using the similarity matrix defined by \ref{simi_Sibo}, we will obtain $S_{1,2}=S_{1,3}=S_{2,3}$ and thus the similarity matrix is not able to separate nodes 1 and 2 from node 3. 

Figure \ref{fig:example1} Example 2 shows another disadvantage of this similarity matrix. One can see that node 1 and node 2 should belong to one same cluster because they have same parents and children. Having a small number of parents and children for both nodes, the value of $S_{1,2}$ will be relatively small so it can be ignored when implementing  low rank projection.

The main idea of the new method is to check the percentage of common "parents" or "children" of two nodes. We will thus make the total number of "parents" or "children" for nodes $i$ and $j$ appear in the denominator. For instance, for two nodes $i$ and $j$, the number of common "children" is given by $(A \cdot A^{T})_{i,j}$). The common percentage of "children" for nodes $i$ and $j$ can thus be expressed by:
\begin{equation}
\dfrac{AA^{T}}{\sqrt{\sum\limits_{k=1}^n A_{i,k}}\sqrt{\sum\limits_{k=1}^n A_{j,k}}} 
\end{equation}
The percentage of common "parents" can be expressed in an analogous way. Finally the similarity matrix can be written as:
\begin{equation}
S_{new_{(i,j)}}=\dfrac{(AA^{T})_{i,j}}{\sqrt{\sum\limits_{k=1}^n A_{i,k}}\sqrt{\sum\limits_{k=1}^n A_{j,k}}} +\dfrac{(A^{T}A)_{i,j}}{\sqrt{\sum\limits_{k=1}^n A_{k,i}}\sqrt{\sum\limits_{k=1}^n A_{k,j}}}
\end{equation}
To compute the new similarity matrix, first normalise each row/column of the adjacency matrix :
\begin{align}
C(i,:)=\dfrac{A(i,:)}{\sqrt{\sum\limits_{k=1}^n A_{i,k}}} &&
D(:,j)=\dfrac{A(:,j)}{\sqrt{\sum\limits_{k=1}^n A_{k,j}}}
\end{align}
The new similarity matrix $S_{new}$ can then be written as
\begin{equation}
S_{new}=[C | D^{T}] \cdot [C | D^{T}]^{T}
\end{equation}
Like Browet's method, to reduce the total complexity of the algorithm, we use a low rank approximation. The $r$-dimensional projection of the similarity matrix,  $S^{r}_{new}$, is given by:
\begin{equation}
S^{r}_{new}=X \cdot X^{T}
\end{equation}
with
\begin{equation}
[C | D^{T}] =U \cdot X_{r} \cdot V
\end{equation}
where $X_{r}$ is a real matrix of size $n\times r$, $U$ and $V$ are orthogonal matrices of size respectively $n\times n $ and $r\times r  $, which means
\begin{equation}
V \cdot V^{T}=I_{r}
\qquad
\text{and}
\qquad
U \cdot U^{T}=I_{n}
\end{equation}
\begin{equation}
X=U \cdot X_{r}
\end{equation}

\begin{figure}[h!]
    \centering
    \includegraphics[scale=0.3]{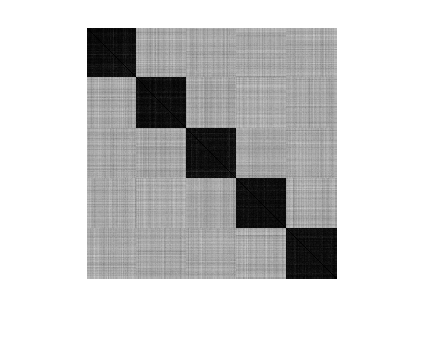}
    \includegraphics[scale=0.3]{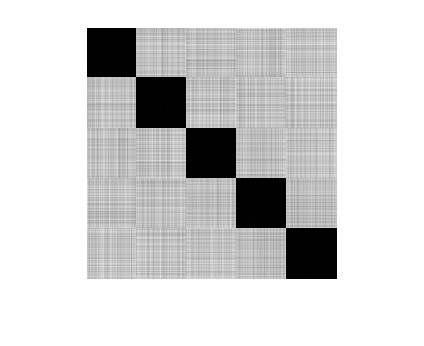}
    \caption{Comparison of similarity matrix using two different similarity measures (Browet's on the left and the new similarity measure on the right) with  $p_{in}=0.9$ and $p_{out}=0.1$ }
    \label{fig:comp_similarity}
\end{figure}

The new similarity takes only one step to compute because we consider only direct connections ("children" and "parents" relationships). It provides a factor matrix which is easier to classify. However, by executing this method, we loose other information in the similarity matrix such as the relationships like "cousin", "nephew" etc. From figure \ref{fig:comp_similarity}, one can see that the similarity matrix obtained by the new measure is actually more "orthogonal" than the one obtained via Browet's method. 

\FloatBarrier

\subsection{Community detection algorithm}
The main objective of the project is to find a permutation matrix $P$ such that $S=(PX)(PX)^{T} $ is as close as possible to a block-diagonal matrix with blocks of ones. This should be done without computing the matrix $S$ directly. Thus, we need to find an efficient way to build a classification of rows of matrix $X$ where similar lines are gathered together.

\subsubsection{Known number of clusters $k$}
\paragraph{$k$-means algorithm}
K-means is a well-known method of vector classification which aims to partition $n$ observations into $k$ clusters where the number of clusters $k$ is given.
Given the lines $x_i$ of the matrix $X$, the $k$-means method assign each of the lines to one of the $k$ clusters $C_j$, with $j = 1,...,k$, to minimise the sum of squared distances of each point in the cluster to the $k$ centres. In other words, the objective is to find
\begin{equation}
\arg \min_{ \{C_j \}_1^k } \sum_{j = 1}^k  \sum_{i \in C_j} \parallel x_i -\mu_{j}\|^{2} 
\end{equation}
where $\mu_{i}$ is the centroid of the cluster $i$ for all $i = 1,...,k$ and $||.||$ is the euclidean norm.\cite{IEEhowto:Kmeans1} Knowing the group to which each line of $X$ should belong, it is then easy to find a permutation such that those lines are grouped together.
\begin{figure}[!h]
    \centering
    \includegraphics[scale=1]{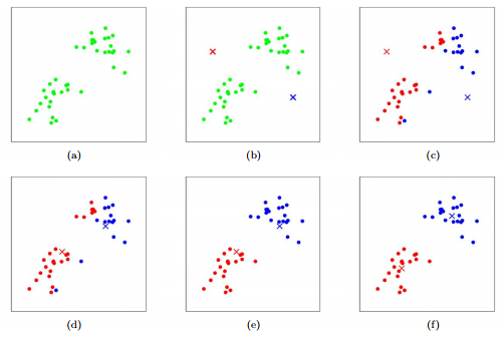}
    \caption{Example of K-means algorithm.ircles., Adaptedfrom website Stanford.edu by  Chris Piech. Based on a handout by Andrew Ng.}
\label{fig_kmeans}
\end{figure}
The goal is thus to regroup $n$ vectors in a $d$-dimensional space to $k$ different clusters.
The algorithm proceeds are as follows:
\begin{itemize}
    \item Start with $k$ initial centres $\mathcal{M} = \{ \mu_1, \mu_2, ..., \mu_k \}$ (not necessarily among the $n$ given points) chosen randomly or intentionally for each cluster (figure \ref{fig_kmeans}(b)).
    \item For $i=1,...,k$, let the cluster $C_i$ be the set of points $x_j \in X$ that are closer to $\mu_i$ than to $\mu_j$ with $i \neq j$ (figure \ref{fig_kmeans}(c).
    \item Once the $n$ elements have been placed into one of the $k$ groups, compute the centroid of each group and replace the $k$ centroids used in the previous step.
    \item Repeat step 2 and 3 until the method converges i.e. when groups don't change any more (figure \ref{fig_kmeans}(e)) or the number of iterations attains the maximum limit.
\end{itemize}
The time complexity of K-means algorithm is given by $\mathcal{O}(I k n t_{dist})$ with $I$ the maximum number of iterations and $t_{dist}$ the time to calculate the distance between two points $x_i$ and $x_j$ of X \cite{IEEhowto:KmeansCompl}.
It should be pointed that the K-means method may converge to different results depending on the initial centres. It is thus of major importance to choose the initial centroids wisely. A good choice may reduce the number of steps needed to get convergence and therefore the total complexity. A bad choice may furthermore lead to arbitrarily bad classification.
\paragraph{Initial guess of $k$-means: $k$-means ++ algorithm} Instead of selecting the initial centres randomly, the k-means ++ algorithm (which gives the default initialisation of the centroids in Matlab) proceeds as follows:
\begin{itemize}
    \item Select an observation uniformly at random from the data set, $X$. The chosen observation is the first centroid, and is denoted $c_1$.
    \item Let $D(x)$ denote the shortest distance from a data point to the closest centre we have already chosen. Choose the new centre $c_i$ to be $x \in X$ with probability $\frac{D(x)^2}{\sum_{x \in X} D(x)^2}$
    \item Repeat step 2 until the $k$ initial centres have been chosen
\end{itemize}
According to Arthur and Vassilvitskii in \cite{IEEhowto:Kmeans++}, k-means++ improves the running time of Lloyd's algorithm, and the quality of the final solution. This randomised greedy strategy takes only $\mathcal{O}(nkd)$ iterations and immediately grants an expected approximation ratio of
$\mathcal{O}(\log(k))$. 

\paragraph{Improving $k$-means algorithm: angle between clusters and co-linearity between elements of the same cluster} To improve the previous algorithm, one might use the properties of the matrix $X$. Two lines of $X$ should have an inner product close to one when the nodes to which they refer belong to the same cluster. They should have an inner product far smaller than one when the nodes belong to different clusters. The value of $S_{i,j}$ is the inner product of $X(i,:)$ and $X(j,:)$. As we have normalised the rows of matrix $X$, the previous result leads to the co-linearity of $X(i,:)$ and $X(j,:)$ when $S_{i,j}=1$.
To check if the classification given by K-means is suitable, we compute the centroid of each group. The inner product of two vectors belonging to different clusters should be smaller than a maximum limit (in most cases, we have chosen this limit as 0.7 by observing the result in figure \ref{fig_innerproduct}). Furthermore, two vectors belonging to the same cluster should have an inner product greater than a minimum limit (we have chosen the minimum value to be 0.9). If the condition is not achieved, K-means will be executed once again until those conditions are satisfied or the maximum number of iterations is reached. If it is the conditions were not satisfied, the algorithm will then just output the result of  the last test by default, leading to a decreasing accuracy.

\newpage
\subsubsection{Computing the number of clusters $k$, unknown}
Until know, the number of clusters $k$ given by the user to the algorithm was supposed to be exact. However, in real life situations when facing graphs with many nodes, represented by huge adjacency matrices, it might be way more difficult to know beforehand the correct number of clusters. In order to find the correct integer $k$, we will present three different methods. In the low rank similarity approximation, Browet uses a projection on dominant subspaces of dimension at most $r$. $r$, which is the dimension of the factor matrix $X$, should be an upper bound of $k$. If the true value of $k$ is higher than $r$ (which means we have under-estimated $r$), it is impossible to make a correct classification on actual matrix $X$. Thus, when the methods fail for all  $k<r$, one should check whether $r$ was not under-estimated. 

\begin{figure}[h]
    \centering
    \includegraphics[scale=0.7]{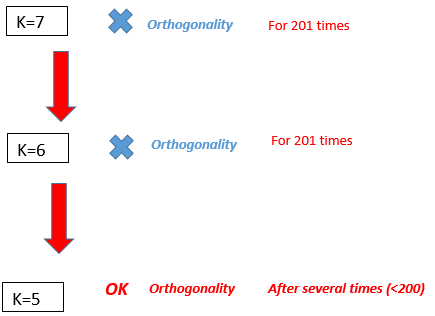}
    \caption{Finding $k=5$ starting from $r=7$ by k-moving method}
    \label{figure_findKHier}
\end{figure}

\paragraph{$k$-moving method} The most natural idea is to try all possible values of $k$ with by increasing or decreasing the initial guess until an acceptable classification is found. Considering $r$ as an upper bound of cluster number $k$, we can simply start with $k=r$. Since we want the algorithm to identify the correct number $k$ automatically, we used the same stopping criterion as in the K-means method using the properties of the matrix $X$. The most useful criterion will be the orthogonality condition since overestimating the value $k$ leads to group elements which indeed belong to the same cluster (and which are thus co-linear) but separates elements which should belong to the same group (and which are thus far from fulling the orthogonality criterion). Those conditions should not be all satisfied when $k$ is not correct. In the case the algorithm finds the correct $k$, it needs $\|r-k \|$ steps to do so. However, it might try to cluster the nodes with the correct $k$ but decide (due to noise or to the complexity of the graph) that it can't find a correct classification with this $k$. Furthermore, this method is more suitable when $k$ is close to $r$. When this is not the case, the complexity will be no longer acceptable.\\
Figure \ref{figure_findKHier} illustrates this method with a simple directed graph of 5 clusters. We can see from this example that at the beginning, we have over-estimated $k$ by taking $r=7$, thus we start testing the classification method when decreasing k by computing at most 201 tests for each value of $k$. Finally, we found the correct number k=5 after testing $k=7$, and $k=6$.

\begin{figure}[!h]
    \centering
    \includegraphics[scale=2]{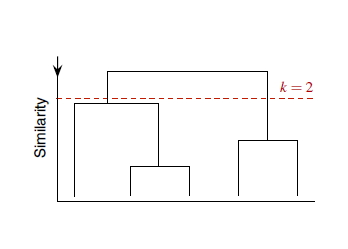}
    \caption{Hierarchical method}
\label{fig_Hierarchicalmethod}
\end{figure}

\paragraph{Hierarchical method} Hierarchical classification method allows to find a correct $k$ with a reduced complexity when $k$ is far from $r$. 
 Given $n$ nodes ${1,2,3....n}$ to be classified, hierarchical clustering proceeds as follows:
\begin{itemize}
    \item Define $n$ groups of singleton :  $G_i = \{i\}$ with $i = 1, ..., n$
    \item Combine two groups which possess the minimum distance between all pairs of groups, i.e.
    \begin{align}
    (k,l)&=\arg\min_{i,j}{d(G_{i},G_{j})} \\
    d(Gi,Gj)&=\parallel \mu_{j} -\mu_{j}\|^{2}
    \end{align}
    with $\mu_{j}$ and $\mu_{i}$  the centroids of the groups $G_{i}$ and $G_{j}$
    \item Take $G_{k}=(G_{k} \cup G_{l})$ and $G_{l}=\phi$
    \item Repeat step 2 and 3 until the minimum distance between two groups is high enough
\end{itemize}
The algorithm should stop when the number of non empty clusters is $k$. An example is given on figure \ref{fig_Hierarchicalmethod} for the case k=2. 
This method has however a complexity  $\mathcal{O}(n^2)$. We can thus not directly apply it on matrix $X$. Given an upper bound $r$ of $k$, we find $r$ clusters using K-means algorithm and asking only for co-linearity between elements of the same cluster. This gives us $r$ "sub-clusters" where the true clusters consist of several "sub-clusters".
The goal is to find clusters such that elements belonging to two different clusters are orthogonal. We start the hierarchical method by giving it the centroids of the $r$ "sub-clusters". The hierarchical method iterates until the centroids of different clusters are not collinear any more (orthogonality criterion). When the correct number of clusters $k$ has been found, we apply the $k$-means method with the found number of clusters $k$. 
\begin{figure}[!h]
    \centering
    \includegraphics[scale=1.4]{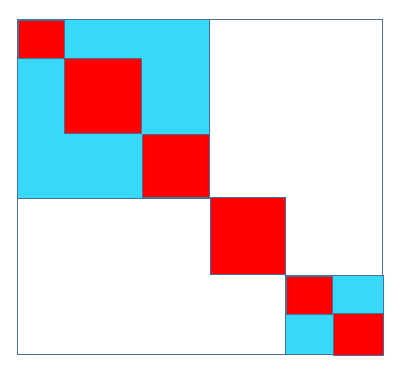}
    \caption{Finding $k=3$ starting from $r=6$ by hierarchical method}
    \label{figure_findKHierB}
\end{figure}
Look for instance at the figure \ref{figure_findKHierB}, where $k=3$ and $r=6$. We start the  preliminary classification by applying $K$-means method to find $6$ "sub-clusters" in red. We then identify that 3 among 6 "sub-cluster" are almost collinear because they are members of one big cluster. We can then regroup another two "sub-clusters" as well. Finally,we find $k=3$ for this directed graph.
Using this method, ideally we will only need to execute K-means method two times.

\paragraph{SVD method} Another way to correctly estimate $k$ when $r > k$ consists in computing the singular value decomposition of a matrix $X$ and compute the number of non-negligible singular values. More formally, for
\begin{align*}
X^{(r)}= U \cdot \Sigma \cdot V &&
U^{T} \cdot U = I_{n} &&
V^{T} \cdot V = I_{r}
\end{align*} 
Where $\Sigma$ is the matrix of singular values,
\begin{equation*}
   \Sigma=
 \begin{pmatrix}
\sigma_{1} & 0 & 0 & 0 & 0& 0\\
0 & \sigma_{2} & 0 & 0 & 0& 0\\
0 & 0 & \ddots & 0 & 0& 0\\
0 & 0 & 0 & \sigma_{r} & 0& 0\\
0 & 0 & 0 & 0 & 0& 0\\
\vdots & \vdots & \vdots & \vdots & \vdots& \ddots\\
\end{pmatrix}
\end{equation*} 
For $r\geq q \geq k $ , if $\sigma_{q}\gg \sigma_{q+1} \geq \sigma_{r} $, we will then start the $k$-means method with $q$ groups. This method is very efficient when there exist a big difference between $r$ and $k$. However, when the level of noise is high, the clustering problem becomes harder: it becomes hard to decide whether a singular value is negligible.\\
For instance, when the reduced graph B, containing 50 elements in each cluster, and the matrix $\Sigma$ are given by :
\begin{align*}
   B=
 \begin{pmatrix}
    0 & 1 & 0\\
    0 & 0 & 1\\
    1 & 0 & 0
  \end{pmatrix}
  &&
  S=
  \begin{pmatrix}
    98.1088 & 0 & 0 & 0 & 0& 0\\
    0 & 62.8004 & 0 & 0 & 0& 0\\
    0 & 0 & 56.0030 & 0 & 0& 0\\
    0 & 0 & 0 & 8.8261 & 0& 0\\
    0 & 0 & 0 & 0 & 8.4482& \ddots\\
    \vdots & \vdots & \vdots & \vdots & \vdots& \ddots\\
   \end{pmatrix}
\end{align*} 
We see that only three singular values of $\Sigma$ are not negligible: $98.1088$, $62.8004$ and $56.003$ and thus set $q$ to $3$. 

\paragraph{Comparison of the different methods} The first method has the highest accuracy among all three methods as we test many possible values for $k$. In reality, it is possible to have several possible classifications (even several possible values of k). Because we have taken both conditions (orthogonality and co-linearity) as a stopping criteria, the algorithm is likely to return an acceptable result. If it comes to $k=0$, it ensures the absence of an acceptable classification. The second and the third methods are more efficient when $r$ is much bigger than $k$. Thus, they are peculiarly useful when we have a vague upper bound of $k$. However, the performance of those two methods can be influenced by a high level of perturbation. 

\section{Results}
\subsection{Measuring the quality of a partition}
To compare the results of each algorithm, we need a quantitative criterion to measure how close the extracted partitions are to the true partition of the benchmark. The most commonly used measures in community detection are based on information theory. As the clustering problem is a generalisation of the community detection problem, it seems fit to use the NMI measure as well. The objective is to quantify how much information about the true partition one can infer from the extracted partition. \\
In information theory, if we us assume $X$ is an event that may occur with a probability $p(x)$, the information contained in the event $X$ is defined by
\begin{equation}
    I(X) = - \log(p(x))
\end{equation}
The entropy is a measure of uncertainty of a random variable. It is defined as the expected information of a single realisation of X, i.e.
\begin{equation}
    H(X) = - \sum_{X = x} p(x) \log(p(x))
\end{equation}
The joint entropy measures the uncertainty of the joint probability distribution $p(x, y)$ to observe $X = x$ and $Y = y$ and is given by
\begin{equation}
    H(X,Y) = - \sum_{X = x, Y = y} p(x,y) \log(p(x,y))
\end{equation}
The mutual information is defined as the shared information between two distributions
\begin{equation}
    I(X,Y) = \sum_{X=x, Y=y} p(x,y) \log \left( \frac{p(x,y)}{p(x)p(y)} \right)
\end{equation}
$I(X,Y) = 0$ when $X$ and $Y$ are totally independent.
In our context, the random variables $X$ and $Y$ correspond to community partitions, so $p(x) = p(X = x)$ is the probability that a node taken at
random belongs to community $x$ in the partition $X$. If $n_x$ is the number of nodes in the community $x$ in the partition $X$ and $n_{xy}$ is the number of nodes that belong to community $x$ in the
partition $X$ and to community $y$ in the partition $Y$, we compute
\begin{align}
    p(x) = \frac{n_x}{n} && p(x,y) = \frac{n_{xy}}{n}
\end{align}
There is no upper bound for $I(X, Y)$, so for easier interpretation and comparisons, a normalised version of the mutual information that ranges from 0 to 1 is desirable (NMI). 
\begin{equation}
    NMI(X,Y) = \frac{I(X,Y)}{\sqrt{H(X) H(Y)}}
\end{equation}\cite{IEEhowto:NMI}

\subsection{Erdos-Renyi graphs}
We applied the pairwise similarity measures to extract roles in Erdos-Renyi graphs containing a prescribed block structure. We first choose the reduced graph $G_B(V_B, E_B)$ where each node represents a role we would like to identify. The random graph $G_A(V_A, E_A)$ is then made by assigning a chosen number of nodes per role, meaning that for each node $i \in V_A$ corresponds a given role $R(i) \in V_B$. We then create the edges $(i,j) \in E_A$ with a probability $p_{in} \in [0,1]$ if there exists a given edge between the corresponding roles in $G_B$, i.e. if $(R(i),R(j)) \in E_B$. If there exist no edges for the corresponding roles in $G_B$, the edge might still be added with a probability $p_{out}$ in $G_A$. If $p_{in}$ is much larger than $p_{out}$, then the role graph $G_B$ is accurately representing the different roles in the graph $G_A$ and the pairwise similarity $S$ between the vertices $V_A$ should allow the extraction of those roles. If $p_{out}$ is much larger than $p_{in}$, then the different roles in $G_A$ are much closely represented by the complement graph of $G_B$ represented by the adjacency matrix $I - G_B$. As the role structure still strongly exists in this complement graph, the similarity measure $S$ should still be able to differentiate them. If probabilities $p_{in}$ and $p_{out}$ are close to each other, the role extraction becomes challenging because the graph becomes closer to a uniform Erdos-Renyi graph with is known to be free of any structure.

\subsubsection{Angle between clusters and co-linearity condition}
\begin{figure}
\centering
\begin{minipage}{.3\textwidth}
  \centering
  \includegraphics[width=1.1\linewidth]{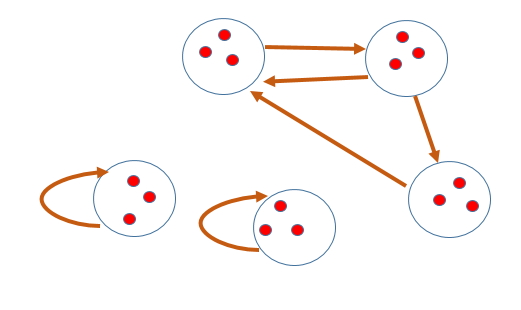}
  \captionof{figure}{Adjacency matrix}
  \label{fig_Adjacencymatrix}
\end{minipage}%
\begin{minipage}{0.3\textwidth}
  \centering
  \includegraphics[width=0.73\linewidth]{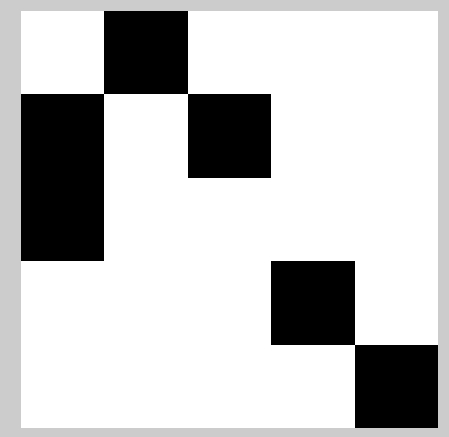}
  \captionof{figure}{Adjacency matrix}
  \label{fig_adjacencymatrix}
\end{minipage}
\begin{minipage}{0.3\textwidth}
  \centering
  \includegraphics[width=0.73\linewidth]{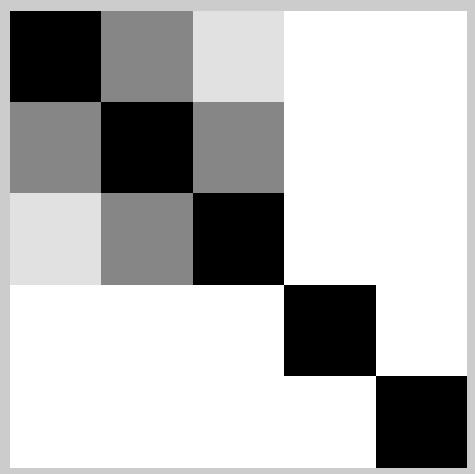}
  \captionof{figure}{Similarity matrix}
  \label{fig_Similaritymatrix}
\end{minipage}
\end{figure}

\begin{figure}[!h]
    \centering
    \includegraphics[scale=1.1]{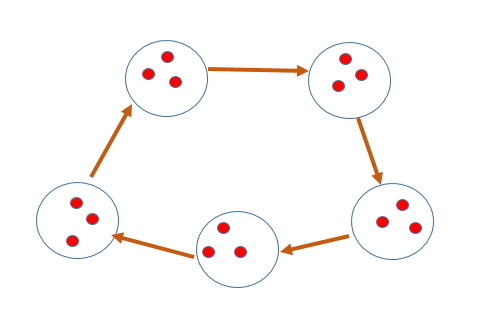}
    \caption{directed graph by cluster}
\label{fig_directedgraph2}
\end{figure}

For instance, choosing a reduced graph of 5 clusters and its adjacency matrix
\begin{equation}
   B=
   \begin{pmatrix}
    0 & 1 & 0 & 0 & 0\\
    1 & 0 & 1 & 0 & 0\\
    1 & 0 & 0 & 0 & 0\\
    0 & 0 & 0 & 1 & 0\\
    0 & 0 & 0 & 0 & 1\\
   \end{pmatrix}
\end{equation}
with 1000 elements in each cluster. The directed graph G and the ideal adjacency matrix A (for which $p_{in} = 1$ and $p_{out} = 0$) associated are thus given by figures  \ref{fig_Adjacencymatrix}. The similarity matrix $S$ defined by Browet is represented on figure \ref{fig_Similaritymatrix}. Values one in the similarity matrix are represented by black pixels and zero values by white ones. We observe from the figure that for two nodes $i,j$ in same cluster $S_{i,j}\simeq 1$ which means they are almost collinear. The white zones outside the diagonal blocks show the orthogonality of two nodes belonging to different clusters. The gray boxes illustrate the fact that elements from different clusters are not always orthogonal to each other, they possess a given angle with respect to each other. In the case of a cyclic graph (shown in the figure \ref{fig_directedgraph2} ), one can see from the similarity matrix that the vectors in different clusters are not all orthogonal to each other.

\begin{figure}[!h]
    \centering
    \includegraphics[scale=0.24]{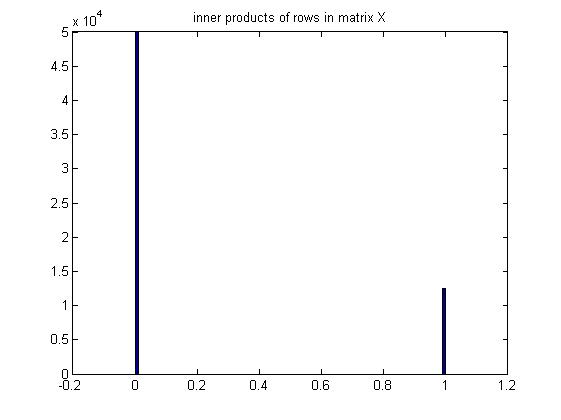}
    \includegraphics[scale=0.24]{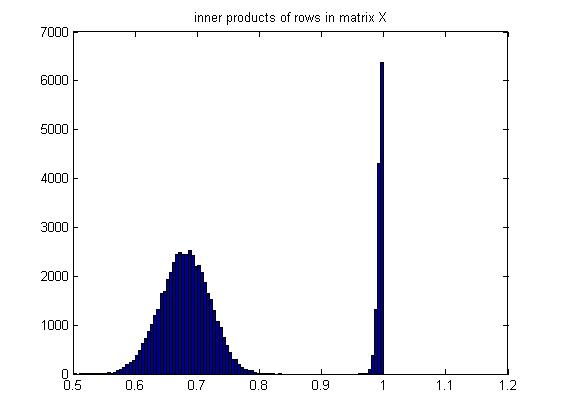}
    \includegraphics[scale=0.24]{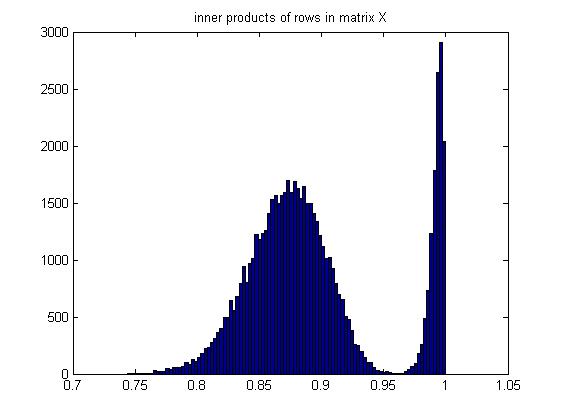}
    \caption{values of inner product with different perturbation level
    from left to right $(p_{in}=1.0,p_{in}=0.0)$,$(p_{in}=0.8,p_{out}=0.2)$,$(p_{in}=0.7,p_{out}=0.3)$}
    \label{fig_innerproduct}
\end{figure}

Indeed, the orthogonality is influenced by the level of noise within the graph. To better detect the influence of noise on the angles between clusters, we choose three different couples $(p_{in},p_{out}) = \{ (1,0), (0.8,0.2), (0.7, 0.2) \}$ and the adjacency matrix of the reduced graph to be
\begin{equation}
   B=
    \begin{pmatrix}
    0 & 1 & 0 & 0 & 0\\
    0 & 0 & 1 & 0 & 0\\
    1 & 0 & 0 & 0 & 0\\
    0 & 0 & 0 & 1 & 0\\
    0 & 0 & 0 & 0 & 1\\
    \end{pmatrix}
\end{equation}
with 200 nodes per block. The histograms of values of the inner product of all pairs of rows in the matrix X are represented in figures \ref{fig_innerproduct}.\\

Analysing the three figures \ref{fig_innerproduct}, we observe two separated peak values of inner products. The first group linked to the first peak has an inner product for from 1, meaning the lines in $X$ are linked to nodes belonging to different clusters. The second group linked to the second peak has an inner product close to one, which means nodes do belong to the same cluster. 
From the same figures, we easily deduce that the amplitude of the perturbation reduces the orthogonality from the vectors (i.e. the minimum angle between two elements from different clusters increase as a function of the perturbation). For instance, in the case $p_{in}=1$ and $p_{out}=0$, there are only two possible values 0 or 1 for inner products. This means vectors in the same cluster are all parallel to each other and two vectors from different clusters are always orthogonal. We can observe also that for $p_{in}$ smaller than 0.7 the set of inner product values is actually dense in the interval [0.6,1]. This will make it more difficult to separate elements from different clusters. The distribution of inner product values depends also on the reduced graph B and number of elements in each cluster, it is easier to separate the clusters when the vectors are "homogeneously" distributed in each group. More examples are shown in the appendix.

\subsubsection{NMI}
\begin{figure}[H]
\centering
\begin{minipage}{.4\textwidth}
  \centering
  \includegraphics[width=0.4\linewidth]{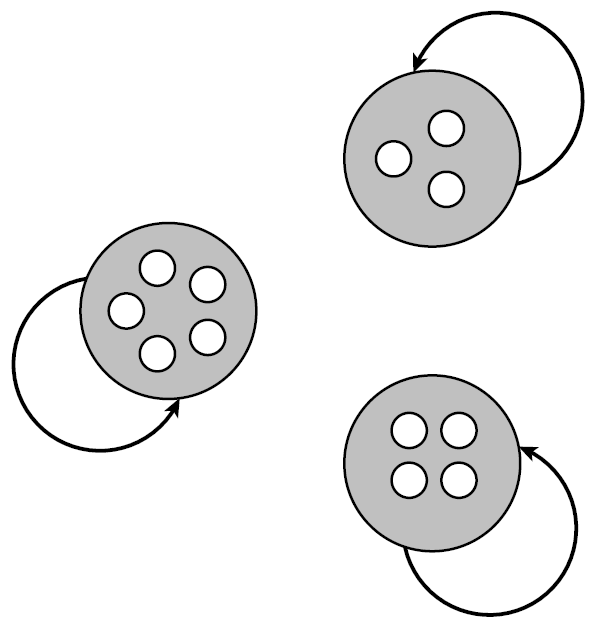}
    \captionof{figure}{Block structure (a), Adapted from “Algorithms for community and role detection in networks,” by A. Browet and P. Van Dooren, 2013}
  \label{fig:BlockA}
\end{minipage}%
\hspace{1.0cm}
\begin{minipage}{0.4\textwidth}
  \centering
  \includegraphics[width=0.4\linewidth]{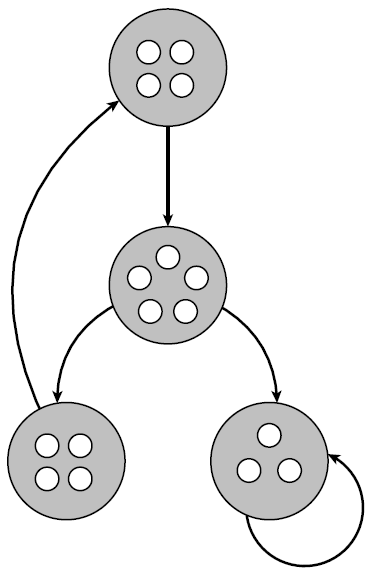}
  \captionof{figure}{Block structure (b), Adapted from “Algorithms for community and role detection in networks,” by A. Browet and P. Van Dooren, 2013}
  \label{fig:BlockB}
\end{minipage}
\end{figure}

\begin{figure}[H]
    \centering
    \includegraphics[width=2.2in]{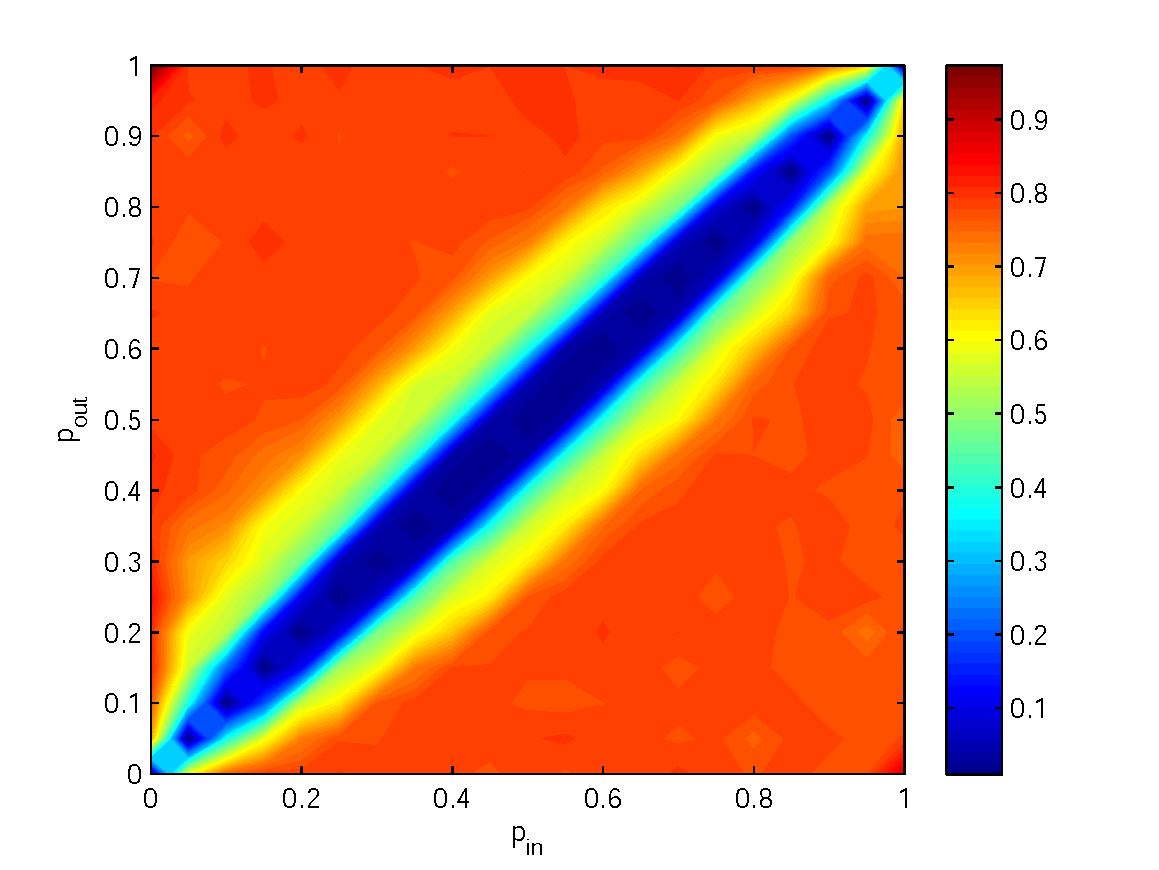}
    \includegraphics[width=2.2in]{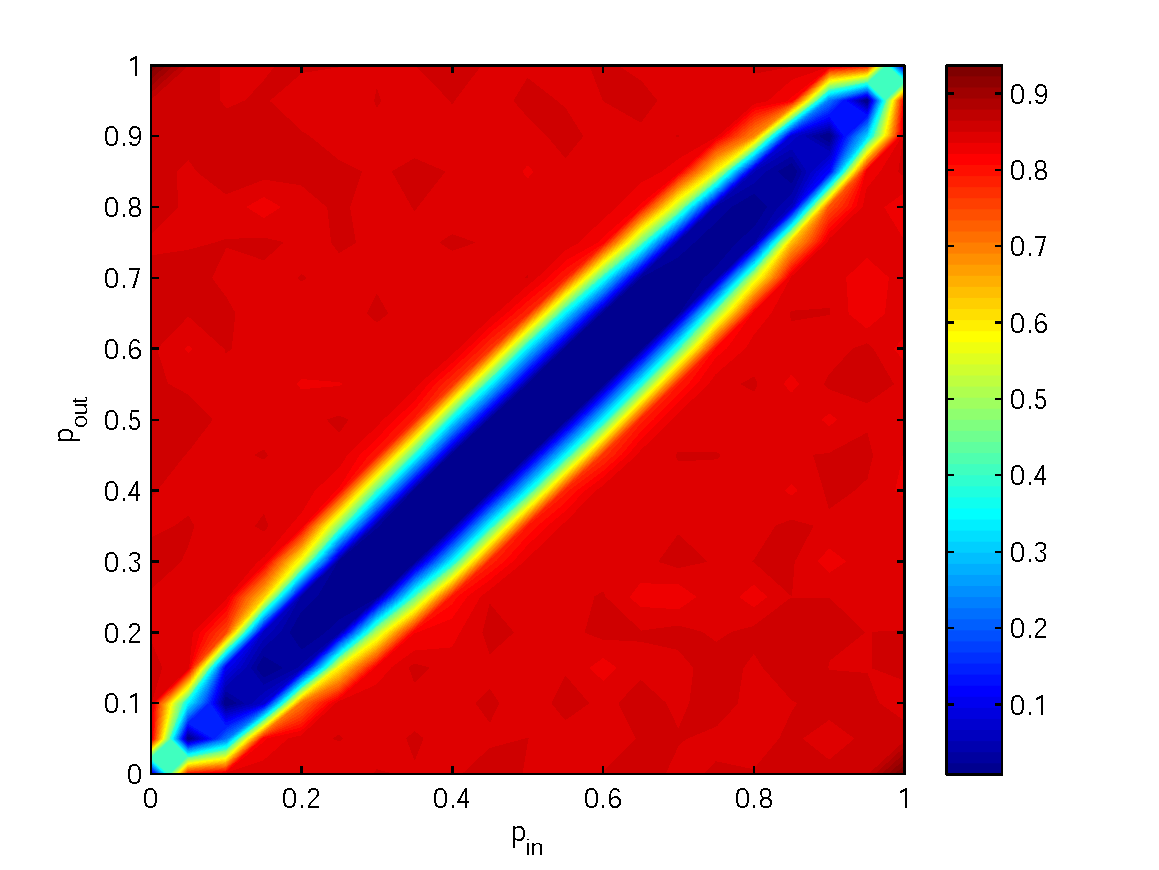}
    \includegraphics[width=2.2in]{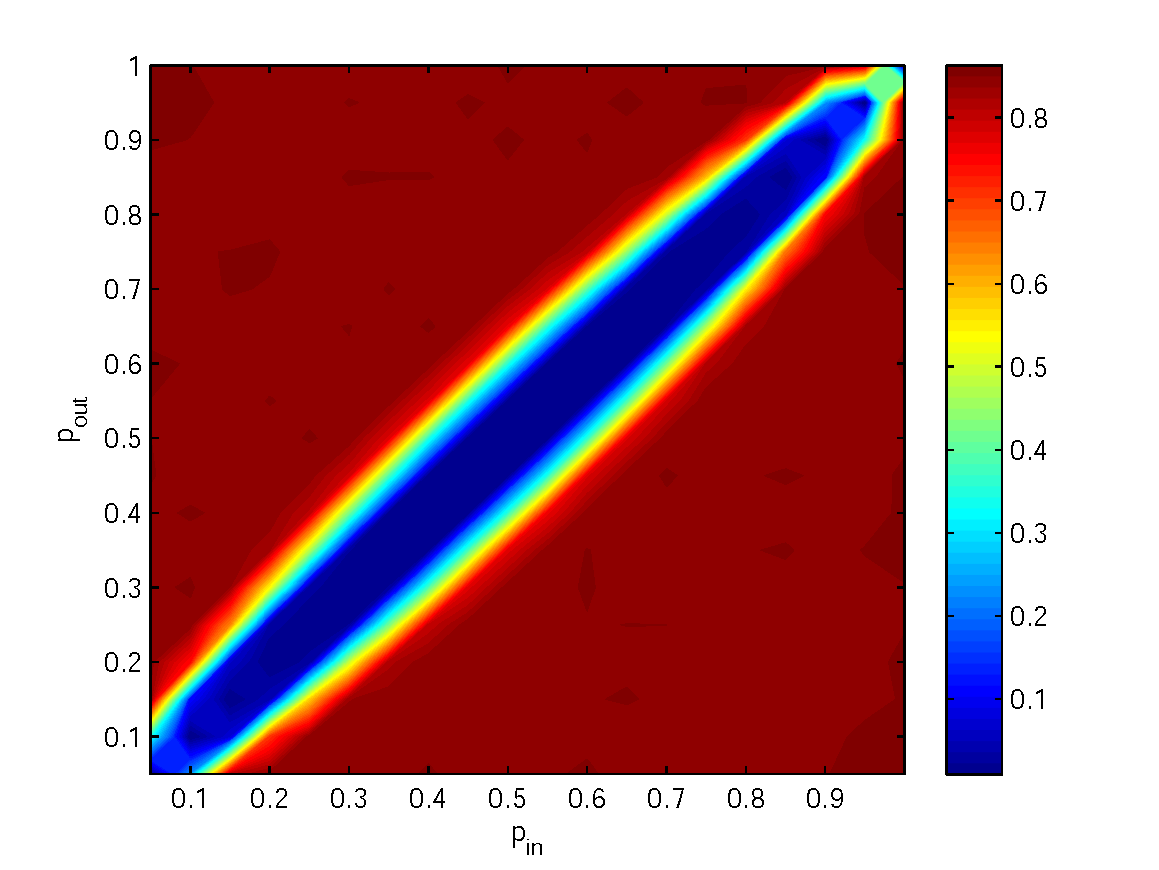}
    \includegraphics[width=2.2in]{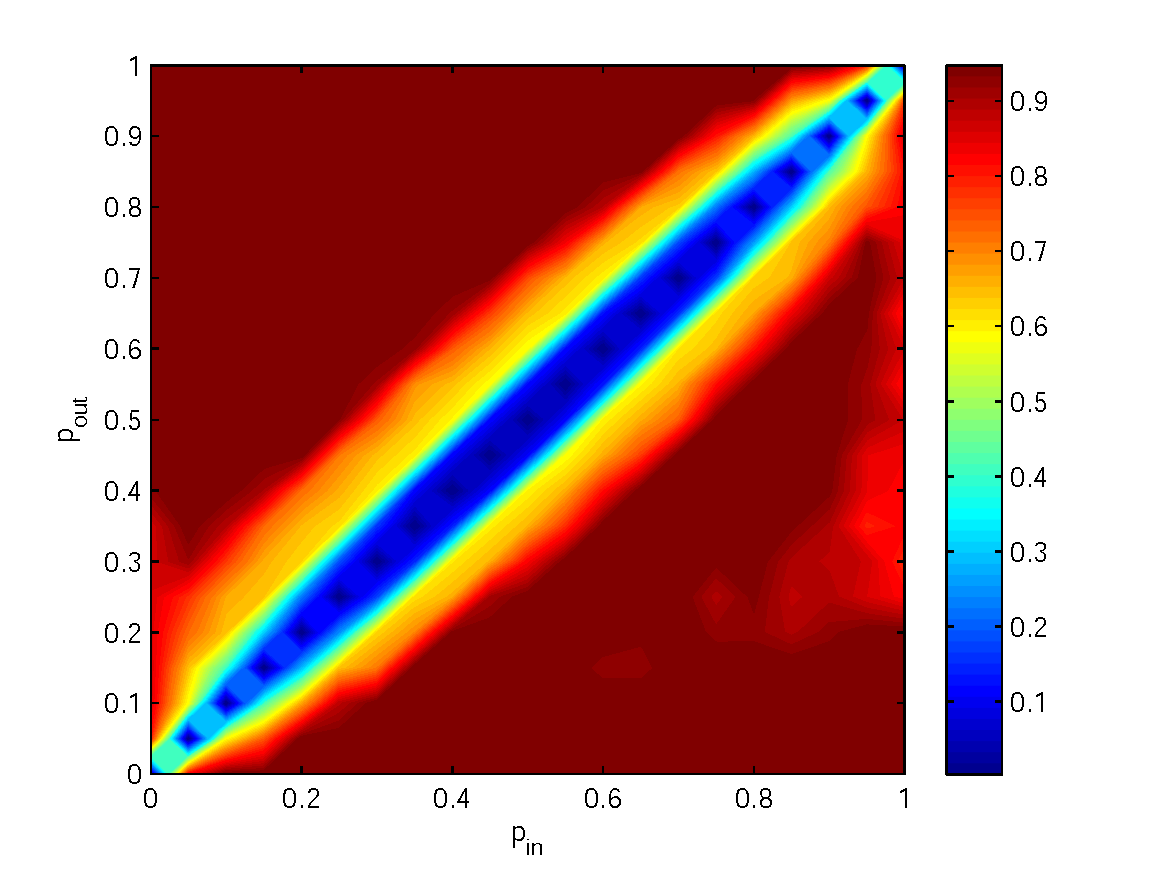}
    \includegraphics[width=2.2in]{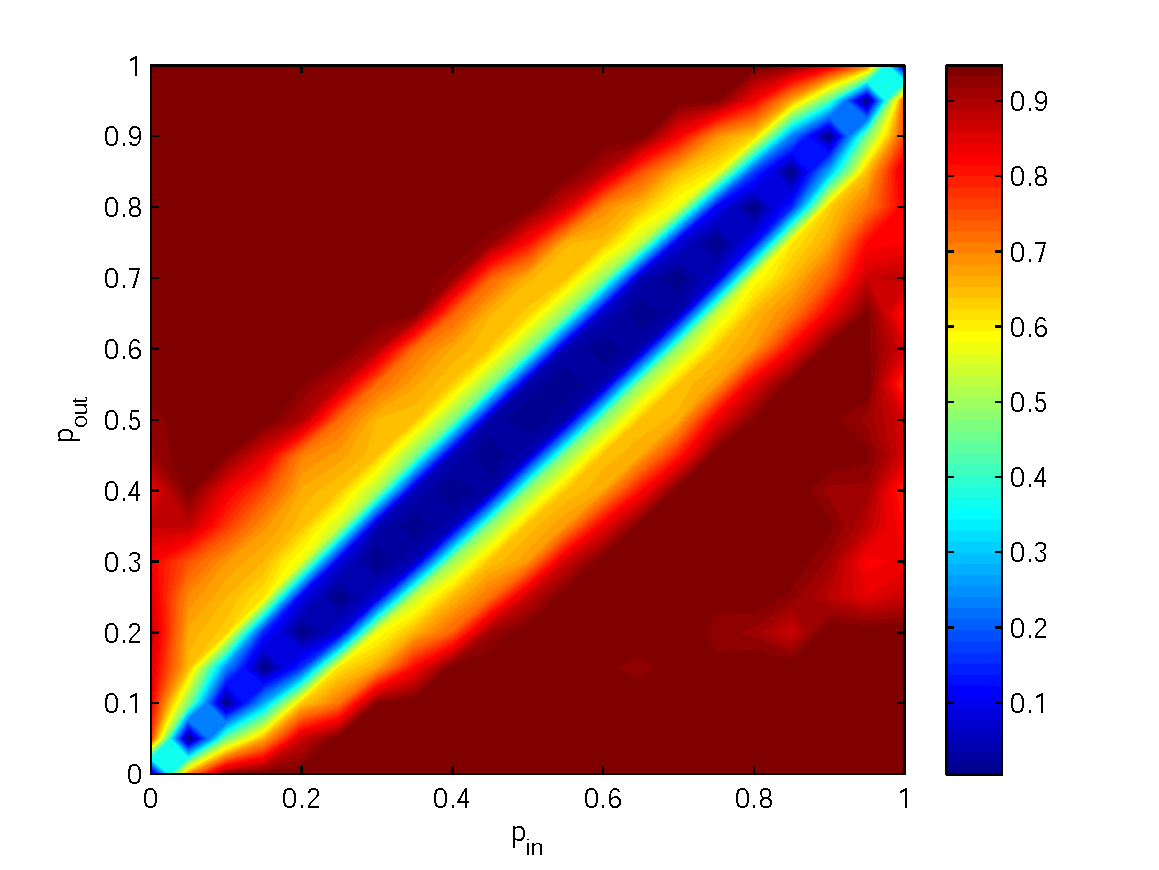}
    \includegraphics[width=2.2in]{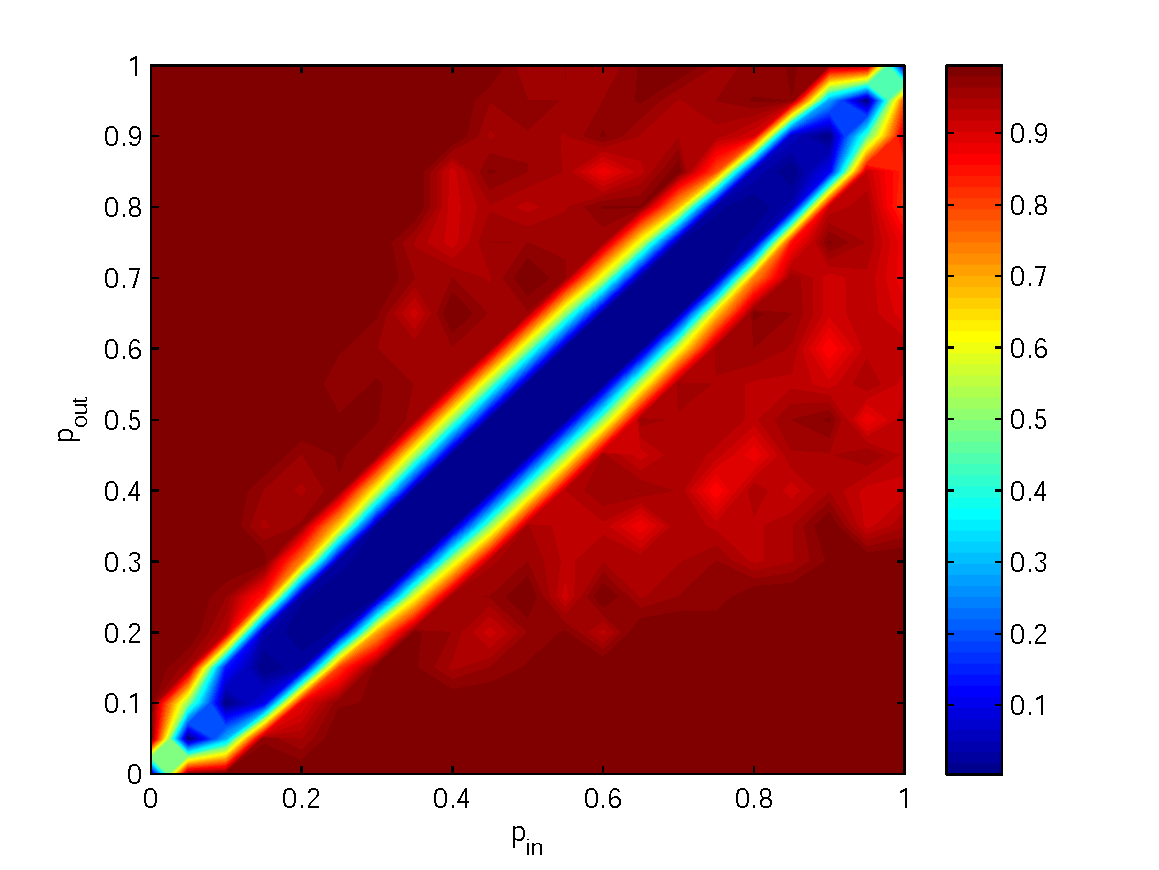}
   	\caption{
    Average normalised mutual information between the exact role structure and the extracted role structure for \ref{fig:BlockA}. \\ 
    First line (from left to right): k-means \\ 
%    \begin{itemize}
%        \item 
$  \star $\hspace{1cm} 20 nodes in the first block and 140 in the two other\\
%        \item  
$ \star $\hspace{1cm} 100 nodes per block\\
%        \item 
$ \star $\hspace{1cm} 100 nodes per block using the new similarity matrix and additional properties of matrix $X$\\
%   \end{itemize}
    Second line (from left to right): k-means using additional properties of the matrix $X$\\
%    \begin{itemize}
%        \item 
$ \star $\hspace{1cm} 30 nodes in the first block and 210 nodes in the other\\
%        \item 
$ \star $\hspace{1cm} 20 nodes in the first block and 140 nodes in the other\\
%        \item 
$ \star $\hspace{1cm} 100 nodes per block\\
%    \end{itemize}
    } 
    %! Argument of \caption@ydblarg has an extra }.<inserted text>\par }
    \label{fig_3Block_Kmeans}
\end{figure}

\begin{figure}[H]
    \centering
    \includegraphics[width=2.2in]{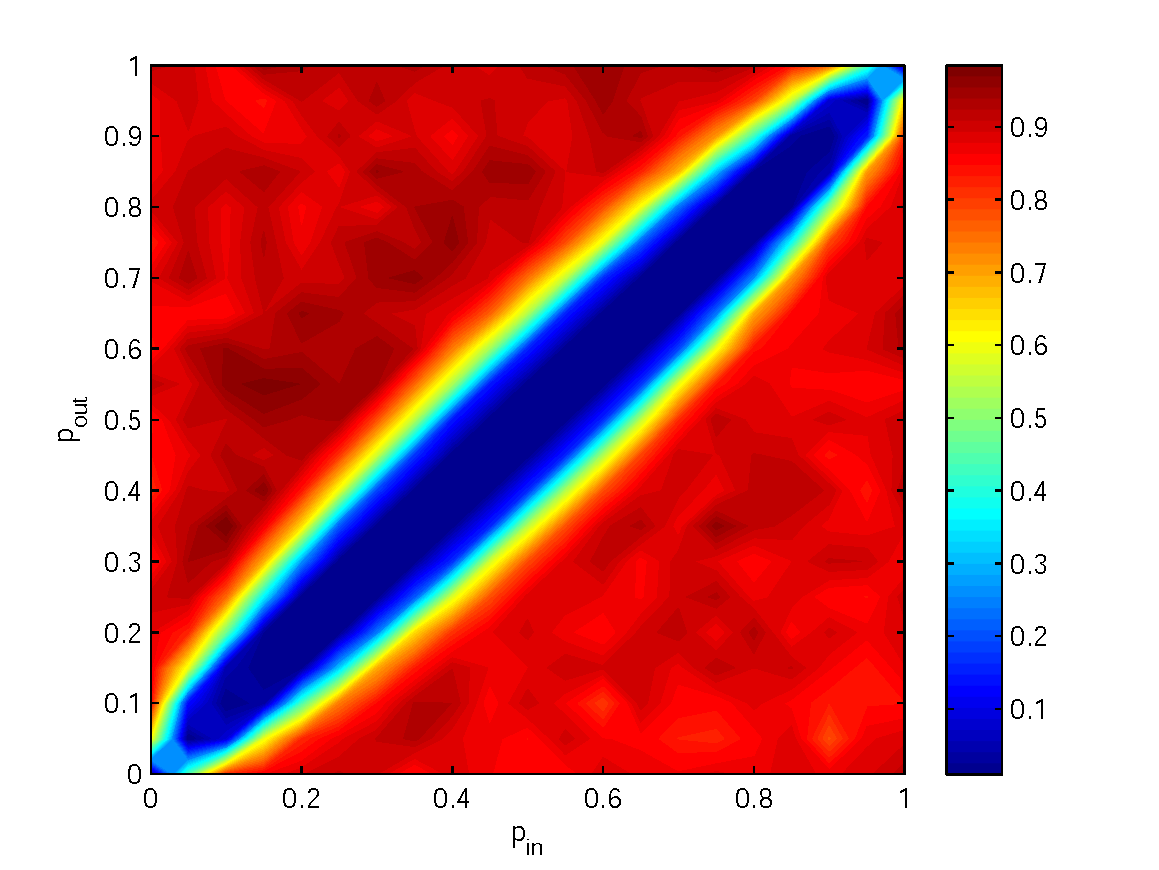}
    \includegraphics[width=2.2in]{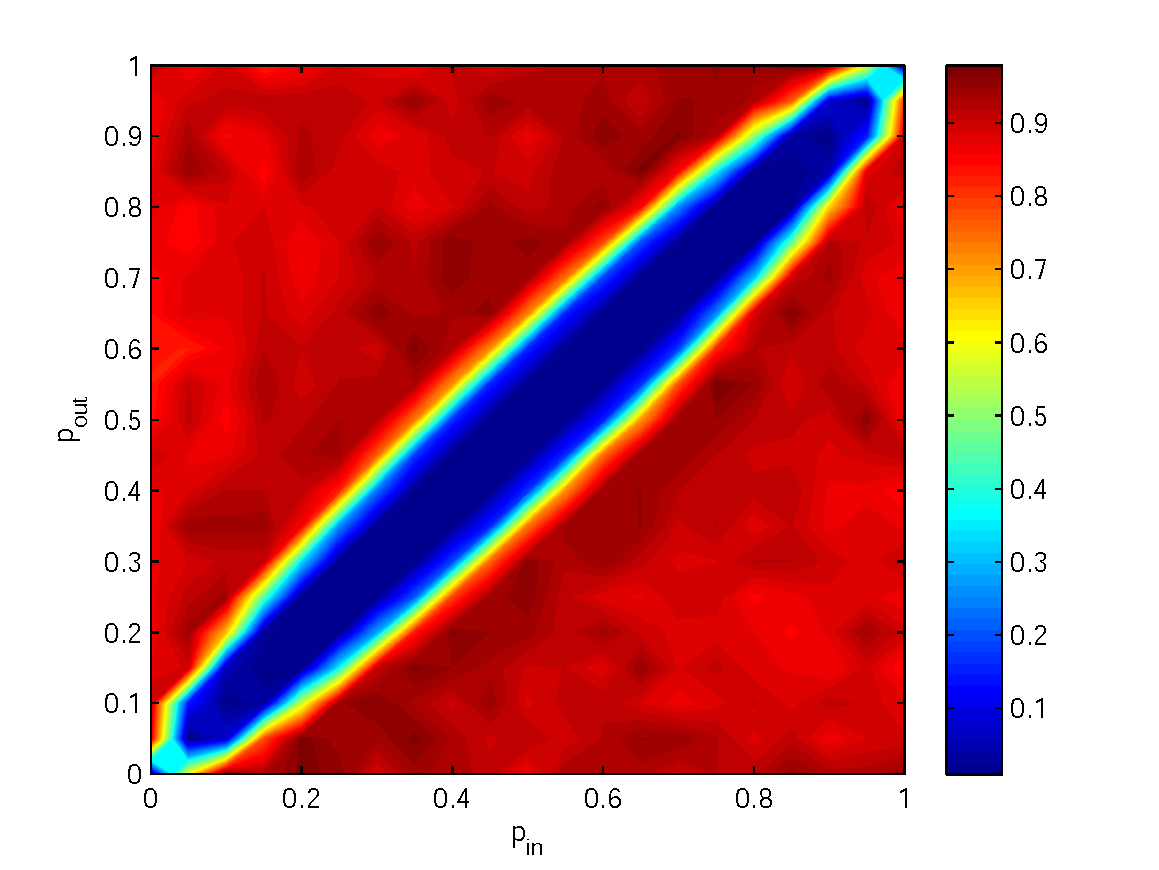}
    \includegraphics[width=2.2in]{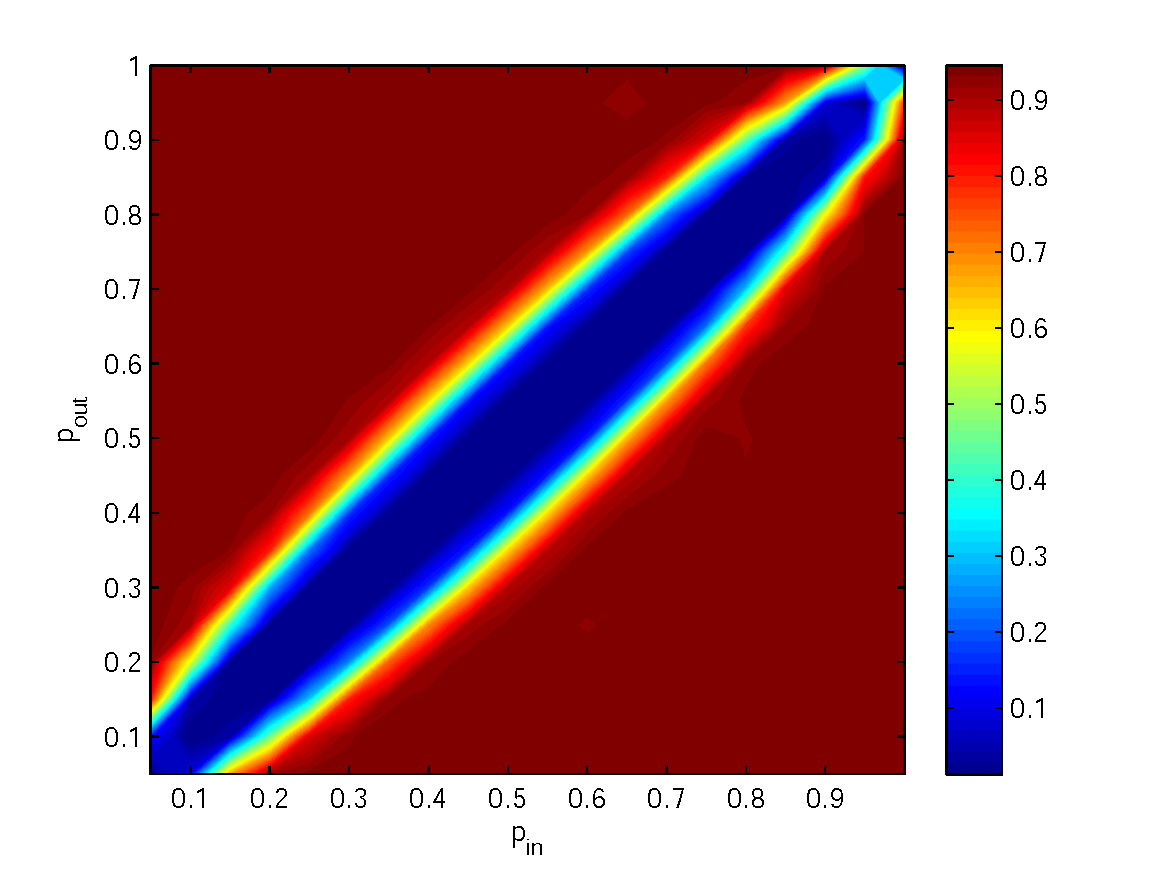}
    \includegraphics[width=2.2in]{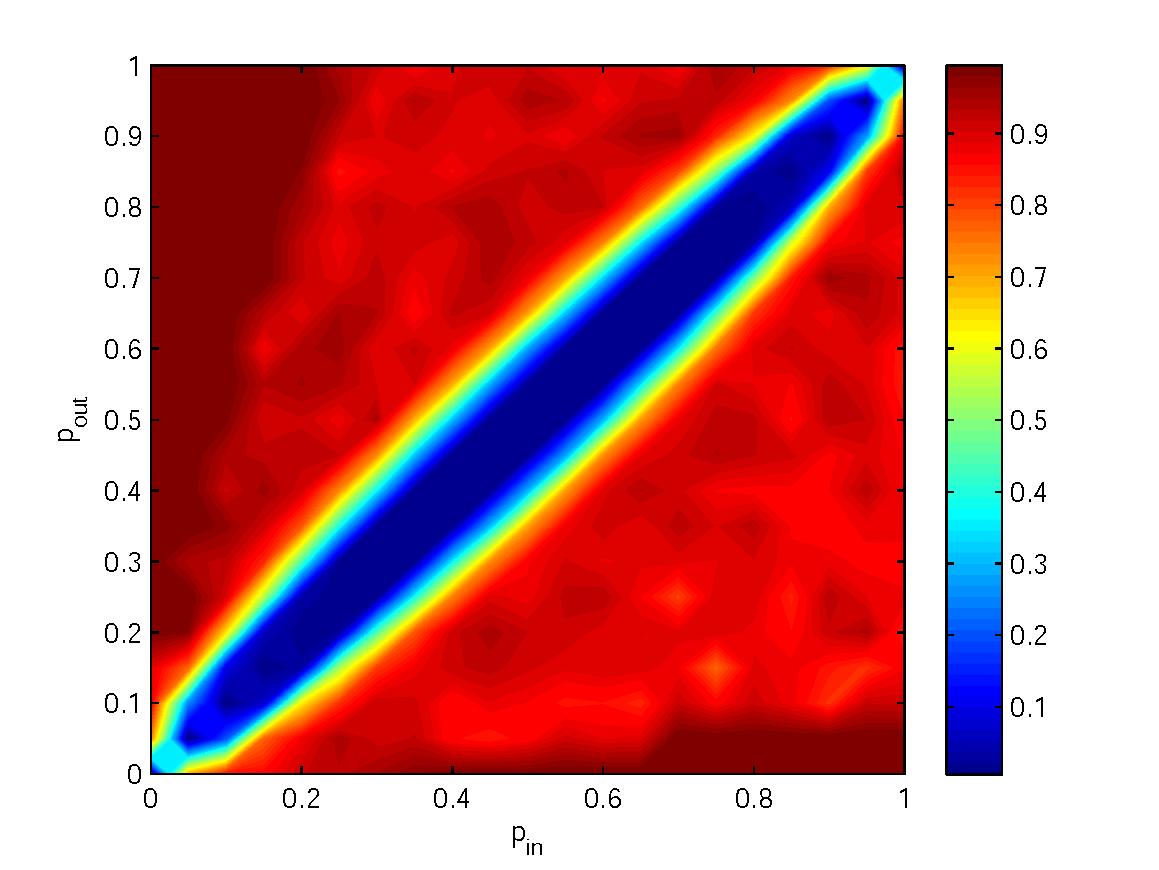}
    \includegraphics[width=2.2in]{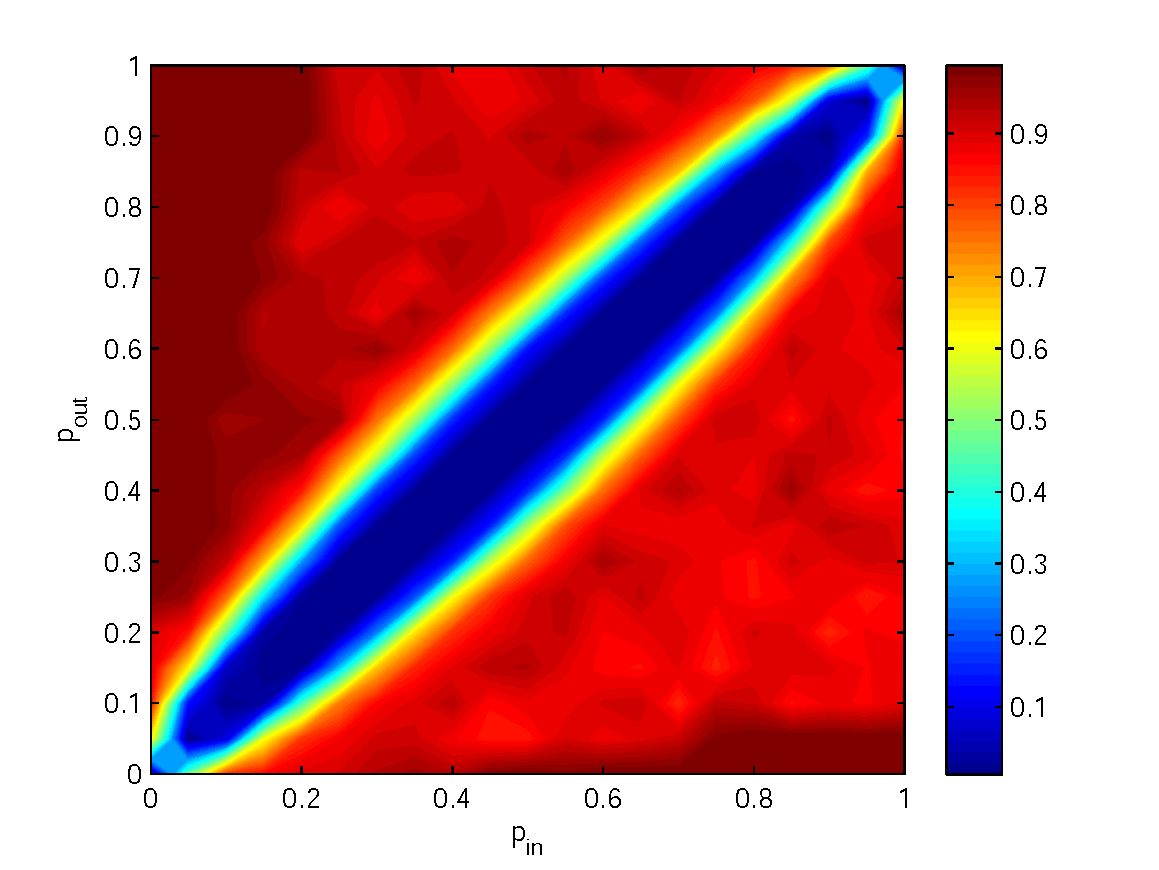}
    \includegraphics[width=2.2in]{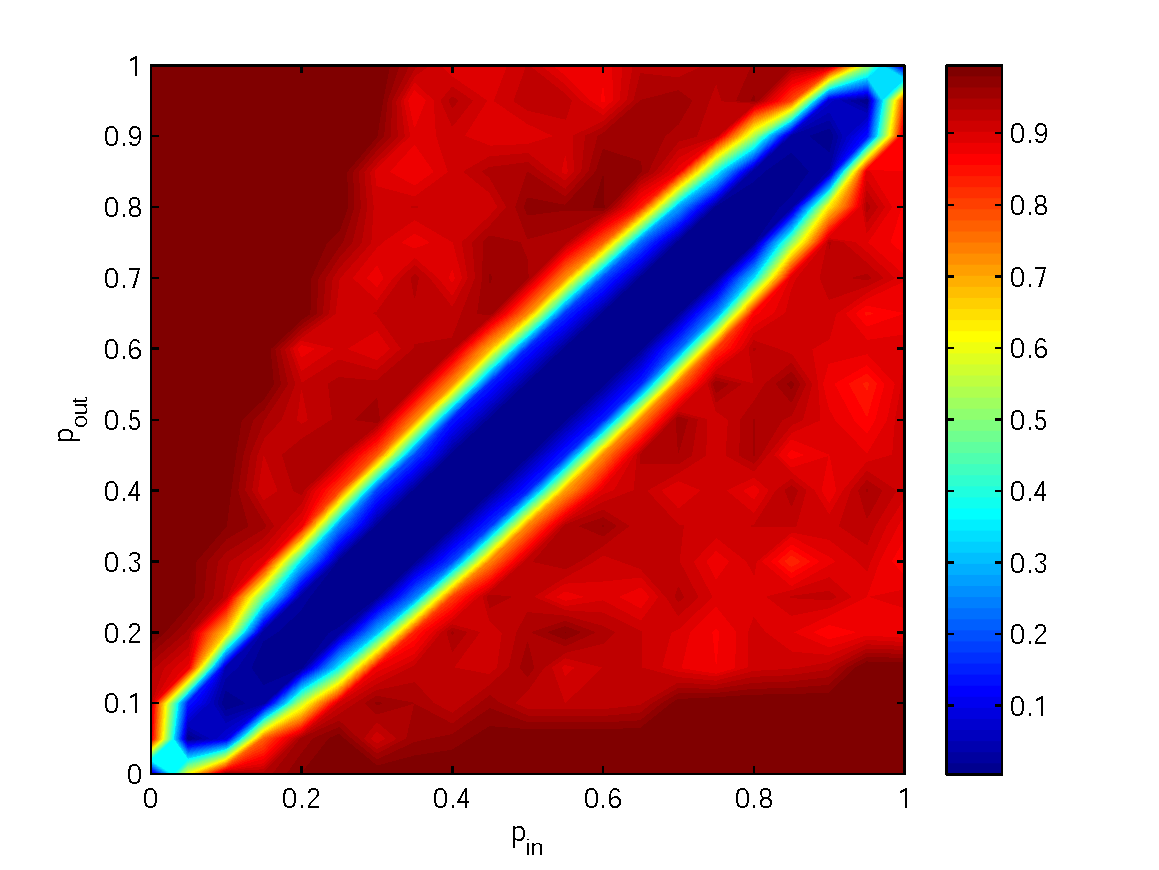}
    \caption{Average normalised mutual information between the exact role structure and the extracted role structure for \ref{fig:BlockB}. \\ 
    First line (from left to right): k-means \\
    %\begin{itemize}
    %    \item 
    $ \star $\hspace{1cm} 10 nodes in the first block and 90 in the three other \\
    %    \item 
    $ \star $\hspace{1cm} 70 nodes per block \\
    %    \item 
    $ \star $\hspace{1cm} 70 nodes per block using the new similarity measure and additional properties of matrix $X$ \\
    %\end{itemize}
    Second line (from left to right): k-means using additional properties of the matrix $X$ \\
    %\begin{itemize}
     %   \item 
     $ \star $\hspace{1cm} 15 nodes in the first block and 135 nodes in the other \\
    %    \item 
    $ \star $\hspace{1cm} 10 nodes in the first block and 90 nodes in the other \\
    %    \item 
    $ \star $\hspace{1cm} 70 nodes per block \\
    %\end{itemize}
    }
    \label{fig_4Block_Kmeans}
\end{figure}

\begin{figure}[H]
    \centering
    \includegraphics[scale=0.4]{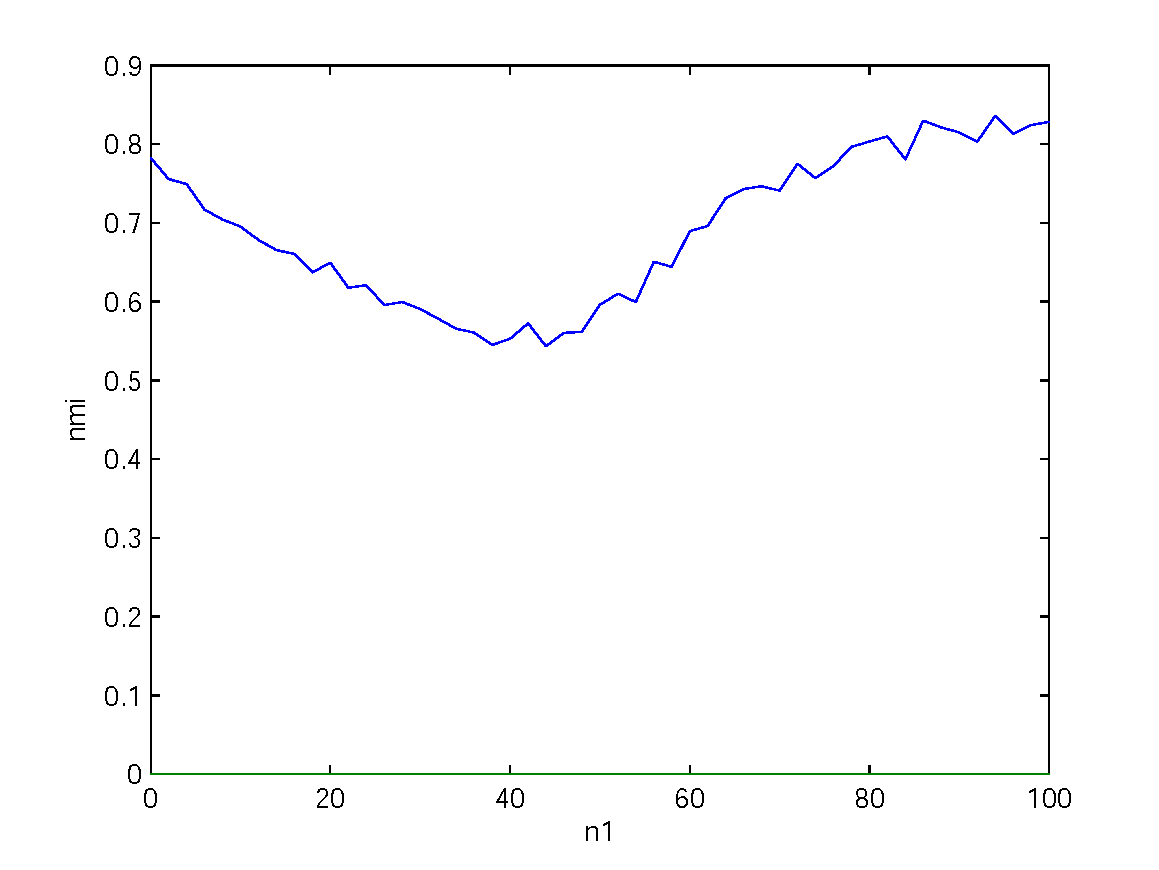}
    \caption{NMI using $k$-means and its additional properties, when varying the size of the first group in figure \ref{fig_3Block_Kmeans}, the total number of nodes kept constant to 300 and the two other groups having the same number of nodes}
    \label{fig:size_FirstGroup}
\end{figure}

We compute the normalised mutual information between the exact role structure and the extracted role partition using the low rank similarity approximation and the similarity measure Salton index method for $r = k$ with $k$ the number of clusters. We generated 20 random realisations for each couple of probability parameters pin and pout in [0, 1] with a step size of 0.05, and we computed the average NMI on those 20 realisations.
\paragraph{Comparison between clustering algorithms}
The community detection algorithm using $k$-means and taking into account the properties of the matrix $X$ performs better than $k$-means alone, as shown on figure \ref{fig_3Block_Kmeans} and figure \ref{fig_4Block_Kmeans}.
\paragraph{Comparison between similarity measures} 
If we compare the results obtained by the two similarity measures, we see the NMI of the new method is more homogeneous when $p_{in}$ is far from $p_{out}$ especially in the structure represented on figure~ \ref{fig:BlockB}. However, even thought this method is faster, the NMI obtained seems slightly worse (especially for the three-blocks structure). 
\paragraph{Effect of size}
When using the same similarity matrix, both community detection algorithms perform better when the number of nodes per group is approximately equal. When the size of one of the groups becomes small compared with the other, the quality of the results decreases up to a certain point when this group becomes too small to strongly impact the NMI. This is shown on figure \ref{fig:size_FirstGroup} where we increased the number of nodes in the first group in the range [0, 100] by step size of 2 while simultaneously decreasing the number of nodes in groups two and three in the range [150, 100] by step size of 1.  Furthermore, the greater the total number of nodes in the graph, the better the performance is when $p_{in}$ is close to $p_{out}$. 

\subsubsection{Time complexity}
\begin{figure}[h!]
    \centering
    \includegraphics[scale=0.5]{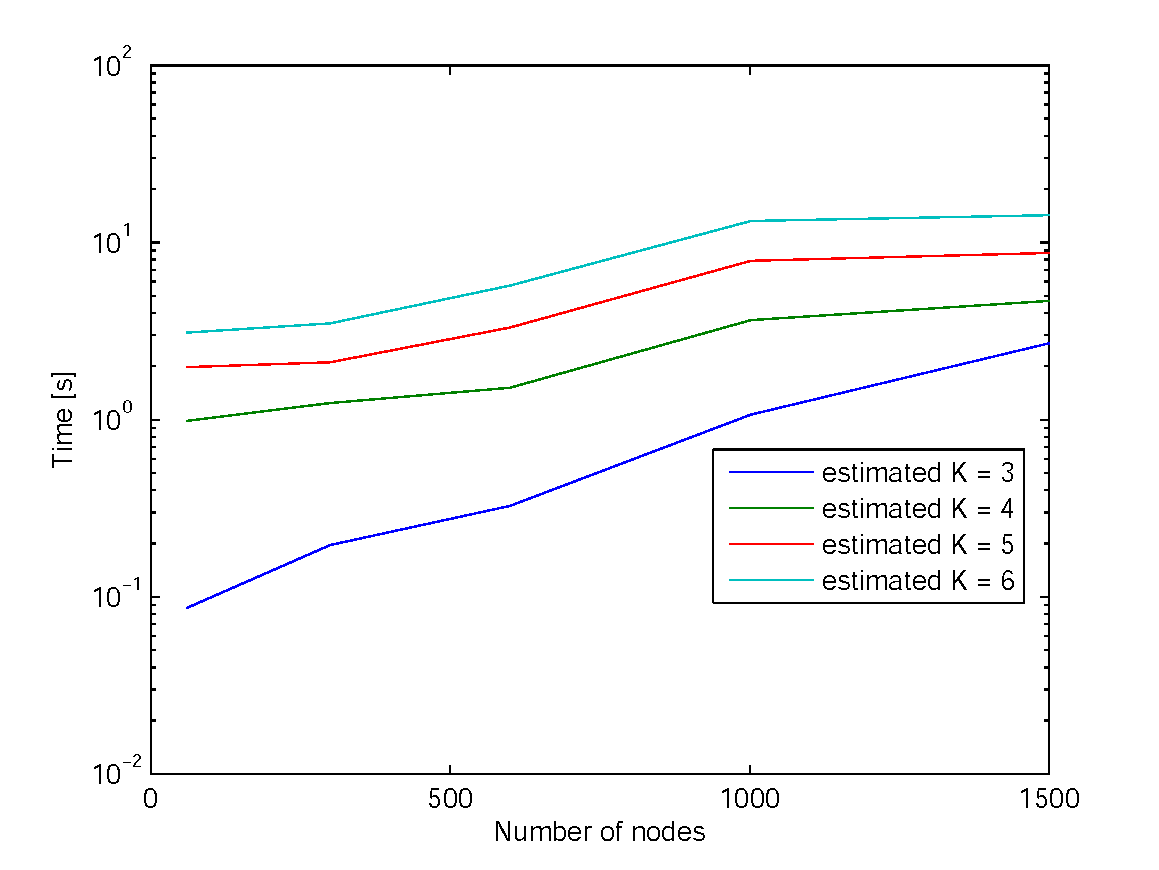}
    \caption{Time complexity when an increasing number of nodes for the correct value of $k=3$ and when detecting the true value of $k$ using $k$-moving algorithm}
    \label{fig:time_kMoving}
\end{figure}

Again, choosing the graph roles as in figure $\ref{fig_3Block_Kmeans}$, we verify on figure $\ref{fig:time_kMoving}$ a linear complexity when the true number of clusters is known. Furthermore, when trying to detect correctly the different clusters and estimating $k$ with the $k$-moving method, the complexity increase with the estimated value for $k$ but remains less than quadratic.

\begin{figure}[!h]
  \centering
  \begin{minipage}[b]{0.4\textwidth}
    \includegraphics[width=\textwidth]{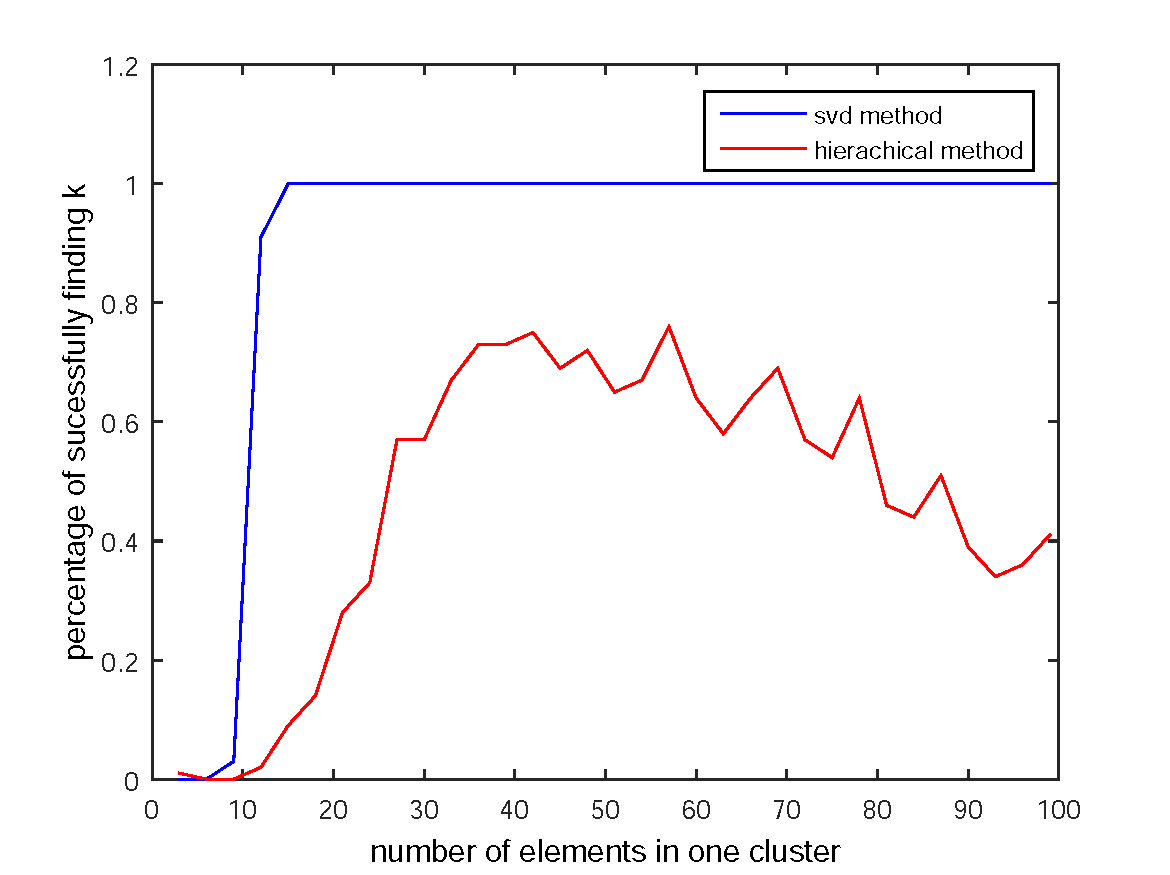}
    \caption{Percentage of correctly chosen $k$}
  \end{minipage}
  \hfill
  \begin{minipage}[b]{0.55\textwidth}
    \includegraphics[width=\textwidth]{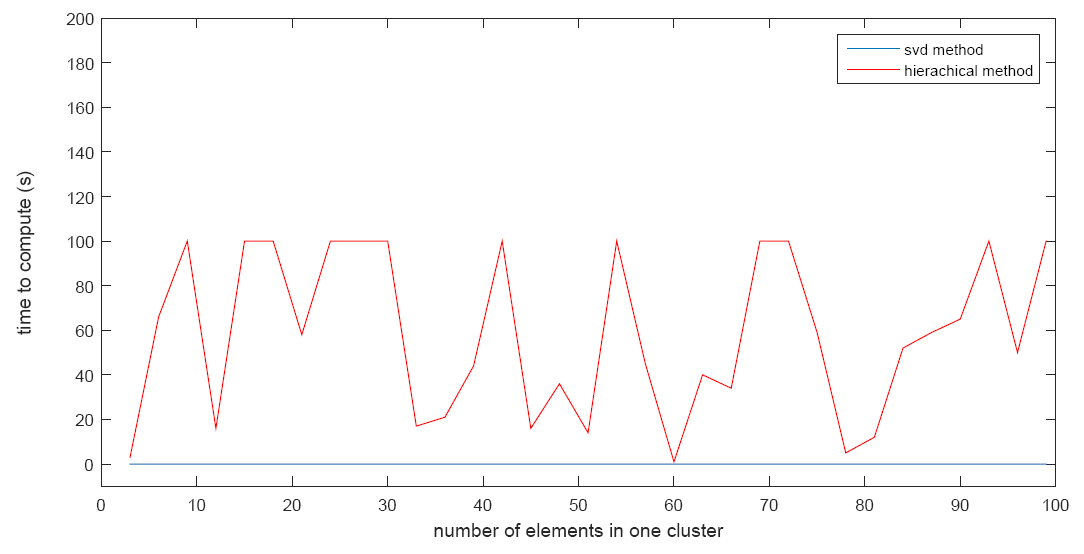}
    \caption{Computation time to find the correct number of clusters}
  \end{minipage}
  \caption{comparison of two methods with high level of perturbation using $p_{in} = 0.7$ and $p_{out}=0.7$}
  \label{figure: comparison}
\end{figure}

Let's now compare the hierarchical method and the singular value method based on their efficiency and accuracy. The figure \ref{figure: comparison} presents the accuracy and time needed for a simple graph of 3 clusters of 200 nodes in total, the number of nodes in the first cluster varying from 0 to 100. The rest of the nodes are distributed homogeneously into two other clusters. One can observe that the SVD method is more accurate than the hierarchical method and costs way less computation time. However, in practice, the gap between the singular values can be much less clear. This may require to ask the users to choose the dimension of projection. 

\begin{figure}[!h]
    \centering
    \includegraphics[scale=0.5]{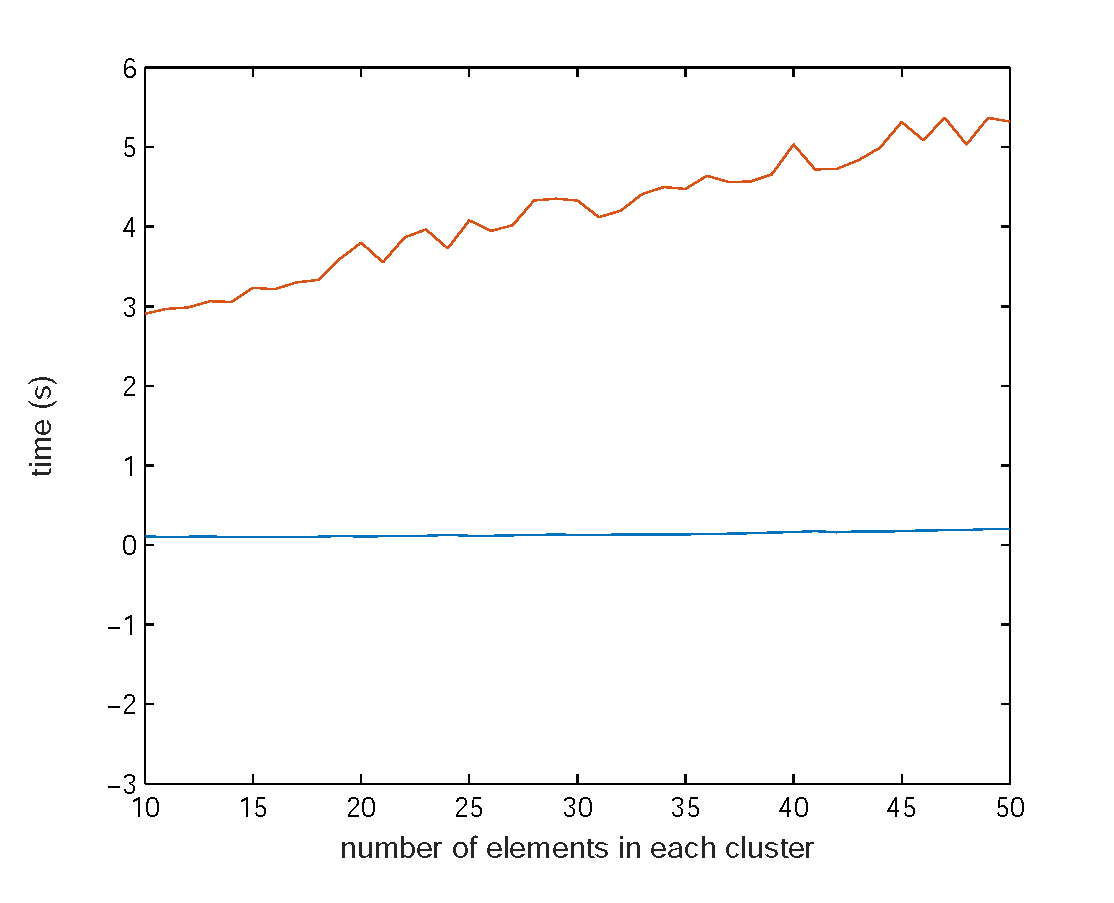}
    \caption{Comparison between computation time for the new simility measure (in blue) and Browet's measure (in red) for $p_{in} = 0.8$ and $p_{out} = 0.2$}
    \label{fig:time_simi}
\end{figure}

As predicted, our method to compute the similarity matrix is far faster than Browet's method (see figure \ref{fig:time_simi}).

\FloatBarrier
\newpage
\subsection{Florida foodweb}
\begin{figure}[!h]
  \centering
  \begin{minipage}[b]{0.35\textwidth}
    \includegraphics[scale=0.3]{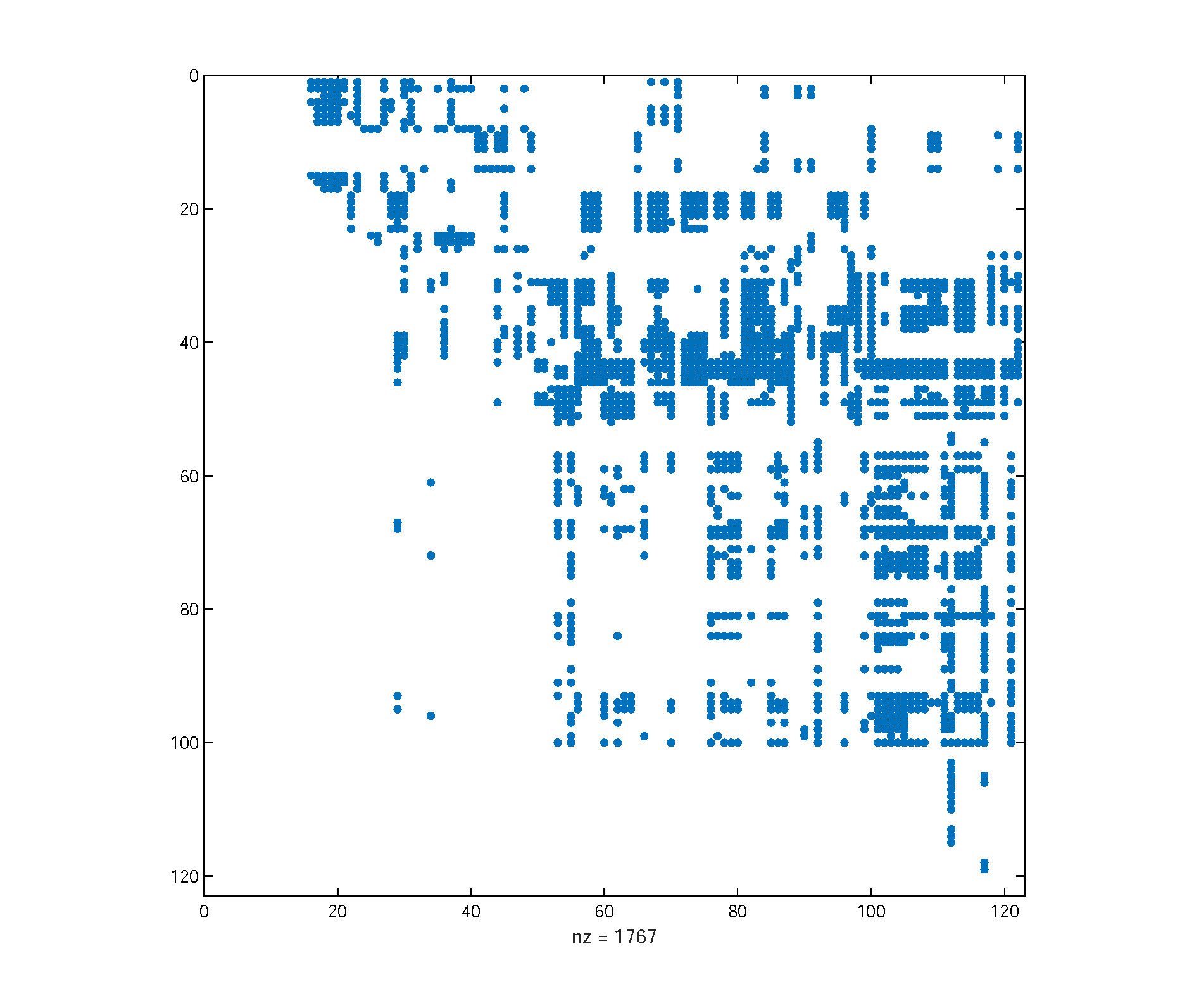}
    \caption{Initial adjacency matrix of the Florida food web without weights.}
    \label{fig:Florida_adj}
  \end{minipage}
  \hfill
  \begin{minipage}[b]{0.35\textwidth}
     \includegraphics[scale=0.5]{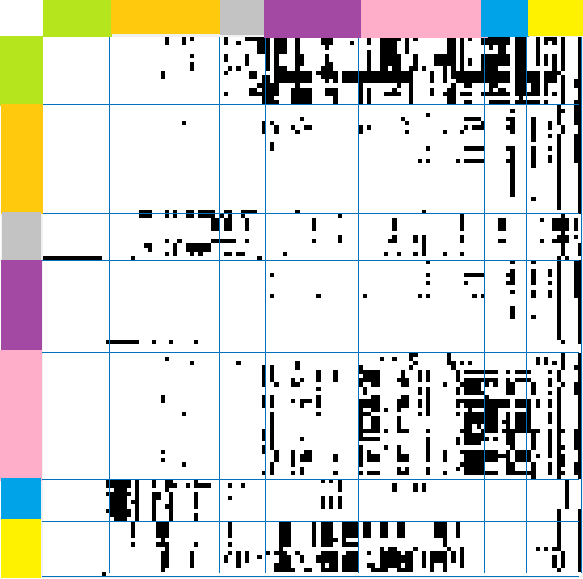}
    \caption{Adjacency matrix of the Florida food web without weights after obtaining clusters}
    \label{fig:Florida_adjB}
  \end{minipage}
\end{figure}

In order to analyse real networks, we analysed the foodweb of the Florida Bay ecosystem \footnote{\url{http://vlado.fmf.uni-lj.si/pub/networks/data/bio/foodweb/foodweb.htm}}. The 122 different biological species can be classified into 7 subgroups: the primary producers (1), the microfauna (2), the macroinvertebrates (3), the fishes (4), the herpetofauna (5), the avifauna (6) and the mammals (7). The Erdos-Renyi graphs for which we validate our algorithm do not have weights on their edges. Therefore, we only considered binary weight on the edges: the element $a_{i,j}$ in the adjacency matrix of the food web has value 1 if the animal $i$ is being eaten by the animal $j$ and 0 otherwise. The initial adjacency matrix is shown on figure \ref{fig:Florida_adj} and its corresponding reduced graph after classified by the biologist on figure \ref{fig_redFloridaBio}. 

\subsubsection{Trying to match biological classification ($k=7$)}
\begin{figure}[!h]
    \centering
    \includegraphics[scale=0.8]{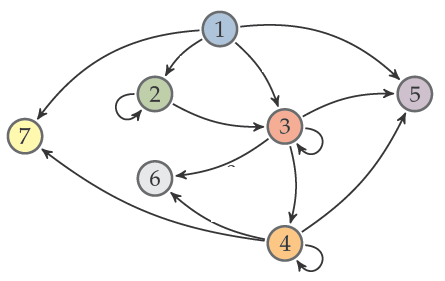}
    \caption{Reduced graph for the natural biological classification of species in Florida bay, Adapted from “Algorithms for community and role detection in networks,” by A. Browet and P. Van Dooren, 2013}
\label{fig_redFloridaBio}
\end{figure}

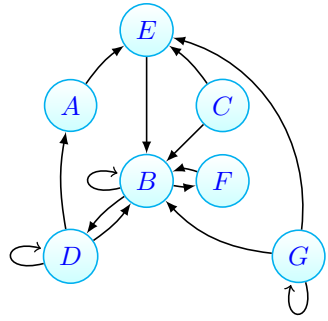
\begin {figure}[!h]
\centering
\begin {tikzpicture}[-latex ,auto ,node distance =1 cm and 1cm ,on grid ,
semithick ,
state/.style ={ circle ,top color =white , bottom color = processblue!20 ,
draw,processblue , text=blue , minimum width =0.5 cm}]
\node[state] (B) {$B$};
\node[state] (A) [above left=of B] {$A$};
\node[state] (C) [above right =of B] {$C$};
\node[state] (D) [below left =of B] {$D$};
\node[state] (E) [above right =of A] {$E$};
\node[state] (F) [below =of C] {$F$};
\node[state] (G) [below right =of F] {$G$};
\path (C) edge [bend right = 10] node[left]{}(E);
\path (C) edge [bend right = 0] node[left]{}(B);
\path(E) edge [bend right = 0] node[left]{}(B);
\path(G) edge [bend right = 40] node[left]{}(E);
\path(G) edge [loop below] node[left]{}(G);
\path(G) edge [bend left = 20] node[left]{}(B);
\path(F) edge [bend right = 20] node[left]{}(B);
\path(D) edge [loop left] node[left]{}(D);
\path(D) edge [bend left = 10] node[left]{}(A);
\path(D) edge [bend right = 10] node[left]{}(B);
\path(A) edge [bend left = 10] node[left]{}(E);
\path(B) edge [bend right = 10] node[left]{}(F);
\path(B) edge [bend right = 10] node[left]{}(D);
\path(B) edge [loop left] node[left]{}(B);
\end{tikzpicture}
\caption{Reduced graph obtained by the clustering algorithm for $k=7$}
\label{fig:Florida_roles}
\end{figure}

The goal is to extract groups of animals sharing the same diet, those groups will be the different clusters. Using the biological classification, the reduced graph of the food web is shown on figure \ref{fig_redFloridaBio}. For clarity, if the proportion of species $i$ in the diet of species $j$ is less than $10 \%$, it is not shown on the graph. \\
After obtaining the clustering of the nodes (for further details see table \ref{tab:classFlorida}), the adjacency matrix after the corresponding permutations is shown on figure \ref{fig:Florida_adjB}. The corresponding reduced graph stands on figure \ref{fig:Florida_roles}.\\

If we analyse table \ref{tab:classFlorida} in detail, we see that Group E includes 71 \% of the macro invertebrates and does not include any other species. The other groups are unfortunately not as uniform. Half of the primary consumers belong to group A while the other half belong to group B. Similarly, if half of the fish are placed in group E, the rest is placed in groups F and G. The first half of the primary consumers is placed in group A and the second half in group B. The herptofauna is splitted in the groups E, F and G while the mammals are placed in groups F and G. 

The difficulty to obtain clear groups can be explained by the fact that some groups like mammals or herptofauna are very small compared to others like fishes. Furthermore, the number of nodes is very small compared to the graphs used before, which increases the difficulty to obtain correct clusters. We could also add that those graphs are very noisy ($p_{in} \approx 0.25$  and $p_out \approx 0.03$) and in those cases, the methods using the property of the graph become useless.

\begin{table}[h!]
    \centering
    \begin{tabular}{|l|p{10cm}|l|}
        \hline
        & Role partition & Biological classification  \\
        \hline
        \cellcolor{airforceblue} A.
        & 2um Spherical Phytoplankt,
         Big Diatoms (>20um),
         Dinoflagellates,
         Drift Algae,
         Oscillatoria,
         Other Phytoplankton,
         Small Diatoms (<20um),
         Synedococcus,
         Syringodium,
         Thalassia&  Primary producer \\
        \cellcolor{airforceblue} 
        & Free Bacteria,
         Benthic Flagellates,
         Water Cilitaes,    
         Water Flagellates  & Microfauna \\
        \hline
        
        \cellcolor{sandstorm} B. & 
        Stone Crab & Macro invertebrates \\
        \cellcolor{sandstorm} & 
        Benthic Ciliates, Benthic Phytoplankton, Drift Algae, Epiphytes, Halodule, Roots, Syringodium, Thalassia, & Primary producer\\
         \cellcolor{sandstorm} & Benthic Flagellates, Meiofauna & Micro fauna \\
        \hline
        
        \cellcolor{celadon} C.
        & Acartia Tonsa,        
        Meroplankton,          
        Oithona nana,         
        Other Copepoda,          
        Other Zooplankton,          
        Paracalanus & Microfauna \\   
        \cellcolor{celadon} & Benthic Crustaceans,
        Detritivorous Amphipods,
        Herbivorous Amphipods,
        Isopods,
        Sponges
        Thor Floridanus            &  Macro inverterbrates \\
        \hline
        
        \cellcolor{coralpink} D.
        & Bivalves,
         Callinectus sapidus,
         Detritivorous Crabs,
         Detritivorous Gastropods,
         Detritivorous Polychaetes,
         Epiphytic Gastropods,
         Herbivorous Shrimp,
         Lobster,
         Macrobenthos,
         Omnivorous Crabs,
         Pink Shrimp,
         Predatory Crabs,
         Predatory Gastropods,
         Predatory Polychaetes,
         Predatory Shrimp,
         Suspension Feeding Polych  & Macroinvertebrates\\
        \hline
        
        \cellcolor{chromeyellow} E. & 
        Anchovy, Bay Anchovy, Blennies, Clown Goby, Code Goby, Dwarf Seahorse, Filefishes, Flatfish, Goldspotted killifish, Grunt, Gulf Pipefish, Halfbeaks, Mojarra, Mullet, Needlefish, Other Cnidaridae, Other Horsefish, Other Killifish, Parrotfish, Pinfish, Rainwater killifish, Sailfin Molly, Sardines, Silverside, Toadfish & Fishes \\
        \cellcolor{chromeyellow} & Echinoderma, Coral, Stone Crab & Macro invertebrates \\
        \cellcolor{chromeyellow} & Green Turtle       & Herptofauna \\
        \hline
        
        \cellcolor{pastelpurple} F. & 
        Bonefish, Catfish, Eels, Lizardfish, Pompano, Porgy, Puffer, Red Drum, Rays, Sharks, Scianids, Snook, Spadefish,  & Fishes \\
        \cellcolor{pastelpurple} & Small Herons \& Egrets, Ibis, Roseate Spoonbill, Herbivorous Ducks, OmnivorousDucks, Gruiformes, Small Shorebirds, Gulls \& Terns & Avifauna \\
        \cellcolor{pastelpurple} & Loggerhead Turtle, Hawksbill Turtle, Brotalus & Herptofauna \\
        \cellcolor{pastelpurple} & Manatee & Mammals \\
        \hline
        
        \cellcolor{timberwolf} G. & 
        Barracuda, Gray Snapper, Grouper, Jacks, Mackerel, Other Pelagic Fishes, Other Snapper, Spotted Seatrout, Tarpon & Fishes \\
        \cellcolor{timberwolf} & Crocodiles & Herpetofauna \\
        \cellcolor{timberwolf} & Big Herons \& Egrets, Comorant, Greeb, Kingfisher, Loon, Pelican, Predatory Ducks, Raptors & Avifauna \\
        \cellcolor{timberwolf} & Dolphin & Mammals\\
        \hline
        
    \end{tabular}
    \caption{Role structure compared to the biological compartments in the Florida Bay network}
    \label{tab:classFlorida}
\end{table}
\FloatBarrier

\subsubsection{Optimal number of groups $k = 3$}
If we use our algorithm to find the optimal number of clusters, we get $k=3$ using WHICH METHOD. The classification then seems more adequate: all primary producers and the micro-fauna are grouped in cluster A; avifauna and mammals are grouped in cluster B, along with big predators; smaller predators are grouped in cluster C (see table \ref{tab:group3Florida}).

\begin{figure}
\centering
\begin{minipage}{.45\linewidth}
  \begin {tikzpicture}[-latex ,auto ,node distance =1.7 cm and 1.7cm ,on grid ,
semithick ,
state/.style ={ circle ,top color =white , bottom color = processblue!20 ,
draw,processblue , text=blue , minimum width =0.6 cm}]
\node[state] (A) {$A$};
\node[state] (C) [above right =of A] {$C$};
\node[state] (B) [below right =of C] {$B$};
\path(C) edge [loop left] node[left]{}(C);
\path(B) edge [loop left] node[left]{}(B);
\path(A) edge [bend right = 0] node[left]{}(C);
\path(C) edge [bend right = 0] node[left]{}(B);
\end{tikzpicture}
    \captionof{figure}{Reduced graph of Florida web obtained by the clustering algorithm for $k=3$}
    \label{fig:Florida_3roles}
\end{minipage}
\hspace{.05\linewidth}
\begin{minipage}{.45\linewidth}
  \includegraphics[width=0.5\linewidth]{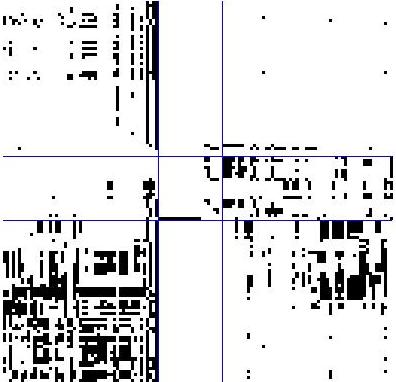}
    \captionof{figure}{Adjacency matrix of Florida web after permutation}
    \label{fig:Florida_3Adj}
\end{minipage}
\end{figure}

To compute the reduced graph of this classification, we made a permutation of the initial adjacency matrix and delimited its different blocks. We then computed the mean number of elements per block. When the mean number of elements in block $(i,j)$ was greater than 0.1, we set an edge between then nodes $i$ and $j$ with $i,j \in \{ A, B, C \}$.
This allows us to conclude from figure \ref{fig:Florida_3roles} animals of group A are being eaten by animals of group C which are themselves eaten by animals of group B. The main downside of this classification is however that animals of group C and group B also eat animals belonging to the same group as themselves but this is also visible in the biological classification. 

\begin{table}[h!]
    \centering
    \begin{tabular}{|l|p{10cm}|l|}
    \hline
    & Role partition & Biological classification  \\
    \hline
     \cellcolor{airforceblue} A &
     2um Spherical Phytoplankt, Synedococcus, Oscillatoria, Small Diatoms (<20um), Big Diatoms (>20um), Dinoflagellates, Other Phytoplankton, Benthic Phytoplankton, Thalassia, Halodule, Syringodium, Roots, Drift Algae, Epiphytes, Free Bacteria, Benthic Ciliates & Primary producer \\
     \cellcolor{airforceblue}  & Water Flagellates, Water Cilitaes, Benthic Flagellates, Meiofauna & Microfauna \\
    \hline
     \cellcolor{coralpink} B & Sharks, Rays, Tarpon, Bonefish, Lizardfish, Catfish, Eels, Needlefish, Snook, Grouper, Jacks, Pompano, Other Snapper, Gray Snapper, Grunt, Porgy, Scianids, Spotted Seatrout, Red Drum, Spadefish, Mackerel, Barracuda, Flatfish, Puffer, Other Pelagic Fishes & Fishes \\         
      \cellcolor{coralpink} & Loon, Pelican, Comorant, Big Herons \& Egrets, Small Herons \& Egrets, Ibis, Roseate Spoonbill, Herbivorous Ducks, Omnivorous Ducks, Predatory Ducks, Raptors, Gruiformes, Small Shorebirds, Gulls \& Terns, Kingfisher & Avifauna \\
      \cellcolor{coralpink} & Crocodiles, Loggerhead Turtle, Hawksbill Turtle, Brotalus  & Herpetofauna \\
     \cellcolor{coralpink} & Dolphin, Manatee & Mammals \\ 
     \hline
     \cellcolor{pastelpurple} C &
     Acartia Tonsa, Oithona nana, Paracalanus, Other Copepoda, Meroplankton, Other Zooplankton & Microfauna \\          
    \cellcolor{pastelpurple} & Sponges, Coral, Echinoderma, Bivalves, Detritivorous Gastropods, Epiphytic Gastropods, Predatory Gastropods, Detritivorous Polychaetes, Predatory Polychaetes, Suspension Feeding Polych, Macrobenthos, Benthic Crustaceans, Detritivorous Amphipods, Herbivorous Amphipods, Isopods, Herbivorous Shrimp, Predatory Shrimp, Pink Shrimp, Thor Floridanus, Lobster, Detritivorous Crabs, Omnivorous Crabs, Predatory Crabs, Callinectus sapidus, Stone Crab,   & Macroinvertebrates  \\ 
    \cellcolor{pastelpurple} & Other Cnidaridae, Sardines, Anchovy, Bay Anchovy, Toadfish, Halfbeaks, Other Killifish, Goldspotted killifish, Rainwater killifish, Sailfin Molly, Silverside, Other Horsefish, Gulf Pipefish, Dwarf Seahorse, Mojarra, Pinfish, Parrotfish, Mullet, Blennies, Code Goby, Clown Goby, Filefishes, Other Demersal Fishes & Fishes \\                
    \cellcolor{pastelpurple} & Green Turtle & Herptofauna \\  
    \hline
    \end{tabular}
    \caption{Role structure for optimal $k = 3$ compared to the biological compartments in the Florida Bay network}
    \label{tab:group3Florida}
\end{table}

\FloatBarrier
\newpage
\subsection{Metal World Trade}
We then classified countries based on trade data from several manufactures of metal among 80 countries in 1993 (Austria, Seychelles, Bangladesh, Croatia, and Barbados), 1994 \footnote{\url{http://vlado.fmf.uni-lj.si/pub/networks/data/esna/metalWT.htm}} and 1995 data (South Africa and Ecuador).  Most missing countries are located in central Africa and the Middle East, or belong to the former USSR. The edges represent imports by one country from another for the class of commodities designated as 'miscellaneous manufactures of metal', which represents high technology products or heavy manufacture. The absolute value of imports (in 1,000 US\$) is used but imports with values less than 1\% of the country's total imports were omitted.

\subsubsection[$k=2$ for Browet's similarity measure]{Trying initial classification with $k=2$, using the additional properties of the graph and Browet's similarity measure}
\begin{figure}[h!]
    \centering
    \includegraphics[scale=0.5]{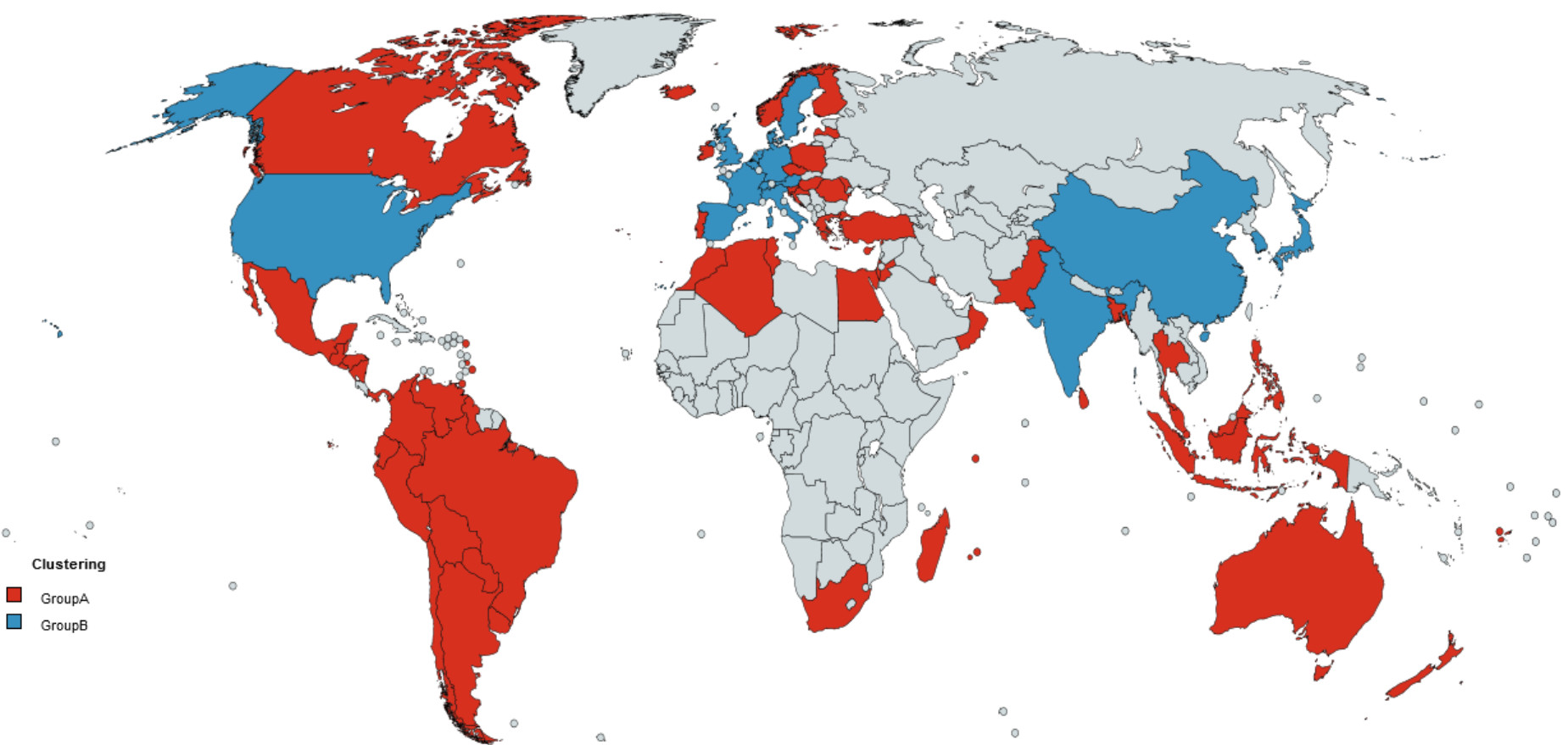}
    \caption{Clustering of countries based on metal trade for $k=2$, made using \url{https://mapchart.net/detworld.html}}
    \label{fig:countries_color2}
\end{figure}

To have a first idea of the interaction between countries, we fix the number of clusters to $k = 2$. We clearly see on figure \ref{fig:countries_color2} that most industrial countries are set in group B and less industrialised countries in group A.

\begin{figure}[h!]
\centering
\begin{minipage}{.45\linewidth}
  \begin {tikzpicture}[-latex ,auto ,node distance =2 cm and 2cm ,on grid ,
semithick ,
state/.style ={ circle ,top color =white , bottom color = processblue!20 ,
draw,processblue , text=blue , minimum width =0.8 cm}]
\node[state] (A) {$A$};
\node[state] (B) [right =of A] {$B$};
\path(B) edge [loop right] node[left]{}(B);
\path(B) edge [bend right = 0] node[left]{}(A);
\end{tikzpicture}
    \captionof{figure}{Reduced graph of metal trade obtained by the clustering algorithm for $k=2$}
    \label{fig:Metal_roles2}
\end{minipage}
\hspace{.05\linewidth}
\begin{minipage}{.45\linewidth}
  \includegraphics[width=0.5\linewidth]{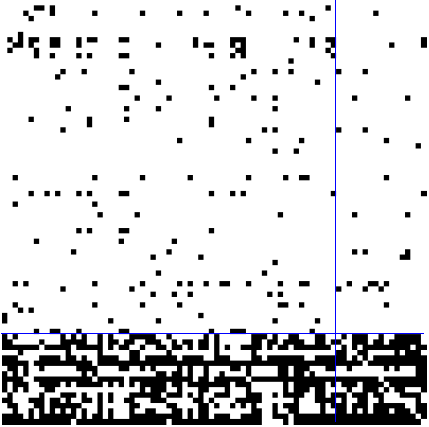}
    \captionof{figure}{Adjacency matrix of countries after permutation}
    \label{fig:countries_adj2}
\end{minipage}
\end{figure}

To compute the reduced graph shown on figure \ref{fig:Metal_roles2}, we computed the adjacency matrix and, separating the different clusters, we computed the mean number of its elements per block. We created an edge $(i,j)$ with $i,j \in \{A, B \}$ when the mean number of elements in the block $i \rightarrow j$ was greater than 0.1. We can then conclude that industrialised countries do trade more than non-industrialised countries, which could have been predicted.

\FloatBarrier
\newpage
\subsubsection{Optimal number of groups $k=4$ using SVD methods, the additional properties of the graph and Browet's similarity measure}

The most industrialised countries are set in group D. The less industrialised countries are set in group A (America) group B (mainly Oceania) and group C (mainly Africa and Eastern Europe). 

\begin{figure}[h!]
\centering
\begin{minipage}{.45\linewidth}
  \begin {tikzpicture}[-latex ,auto ,node distance =1.7 cm and 1.7cm ,on grid ,
semithick ,
state/.style ={ circle ,top color =white , bottom color = processblue!20 ,
draw,processblue , text=blue , minimum width =0.6 cm}]
\node[state] (A) {$A$};
\node[state] (B) [above =of A] {$B$};
\node[state] (C) [right =of A] {$C$};
\node[state] (D) [above =of C] {$D$};
\path(A) edge [loop left] node[left]{}(A);
\path(B) edge [loop left] node[left]{}(B);
\path(D) edge [loop above] node[left]{}(D);
\path(D) edge [bend right = 0] node[left]{}(A);
\path(D) edge [bend right = 0] node[left]{}(B);
\path(D) edge [bend right = 0] node[left]{}(C);
\end{tikzpicture}
    \captionof{figure}{Reduced graph of metal trade obtained by the clustering algorithm for $k=4$}
    \label{fig:Metal_roles}
\end{minipage}
\hspace{.05\linewidth}
\begin{minipage}{.45\linewidth}
  \includegraphics[width=0.5\linewidth]{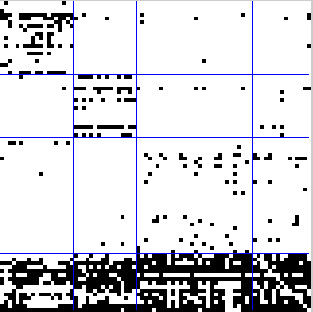}
    \captionof{figure}{Adjacency matrix of countries after permutation}
    \label{fig:countries_adj}
\end{minipage}
\end{figure}

To compute the reduced graph shown on figure \ref{fig:Metal_roles}, did the same as in the previous section and created an edge $(i,j)$ with $i,j \in \{A, B, C, D \}$ when the mean number of elements in the block $i \rightarrow j$ was greater than 0.1. The same conclusions about trade can be made as in the previous section. Furthermore, less industrialised countries do not exchange much with other less industrialised countries located in another geographical area. More industrialised countries possess a higher PIB and can thus afford to trade with countries situated in other geographical areas. This agrees with economical models such as the gravity model of trade (for further details, see \footnote{\url{http://vi.unctad.org/tpa/web/docs/ch3.pdf}}).

\begin{figure}[h!]
    \centering
    \includegraphics[scale=0.4]{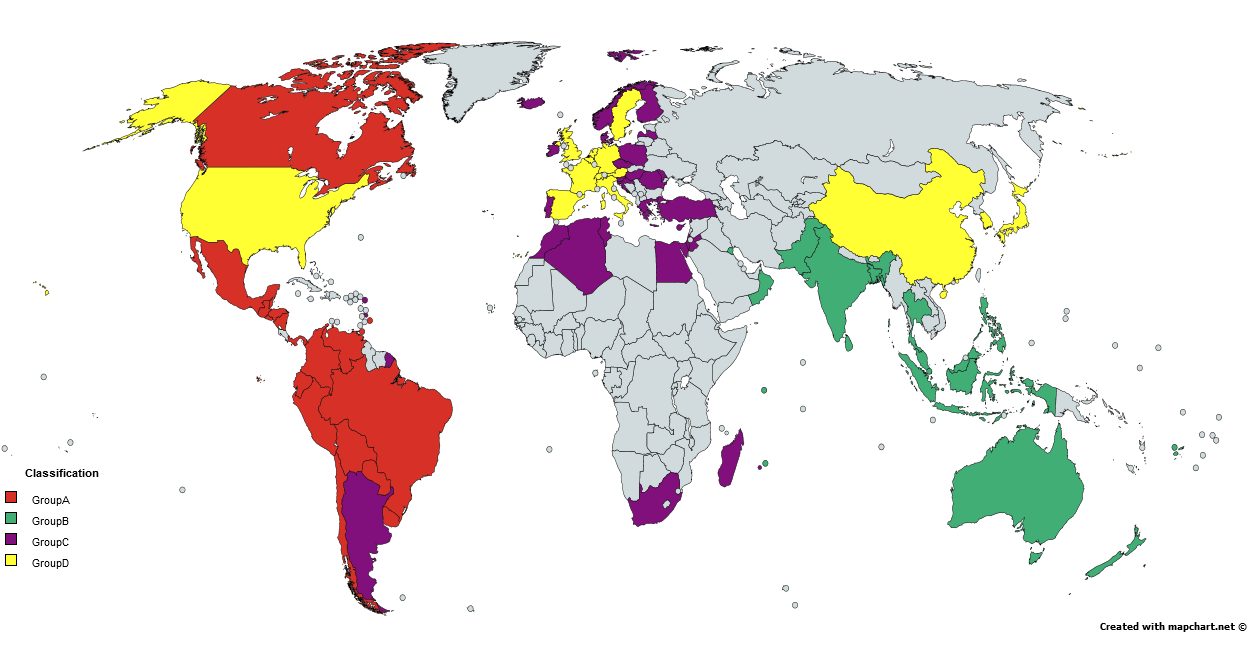}
    \caption{Clustering of countries based on metal trade for $k=4$, made using \url{https://mapchart.net/detworld.html}}
    \label{fig:countries_color}
\end{figure}

\FloatBarrier
\newpage
\subsubsection[$k=4$ using SVD method and new similarity measure]{Optimal number of groups $k=4$ using SVD methods, the additional properties of the graph and the new similarity measure}
\begin{figure}[!h]
    \centering
    \includegraphics[scale=0.5]{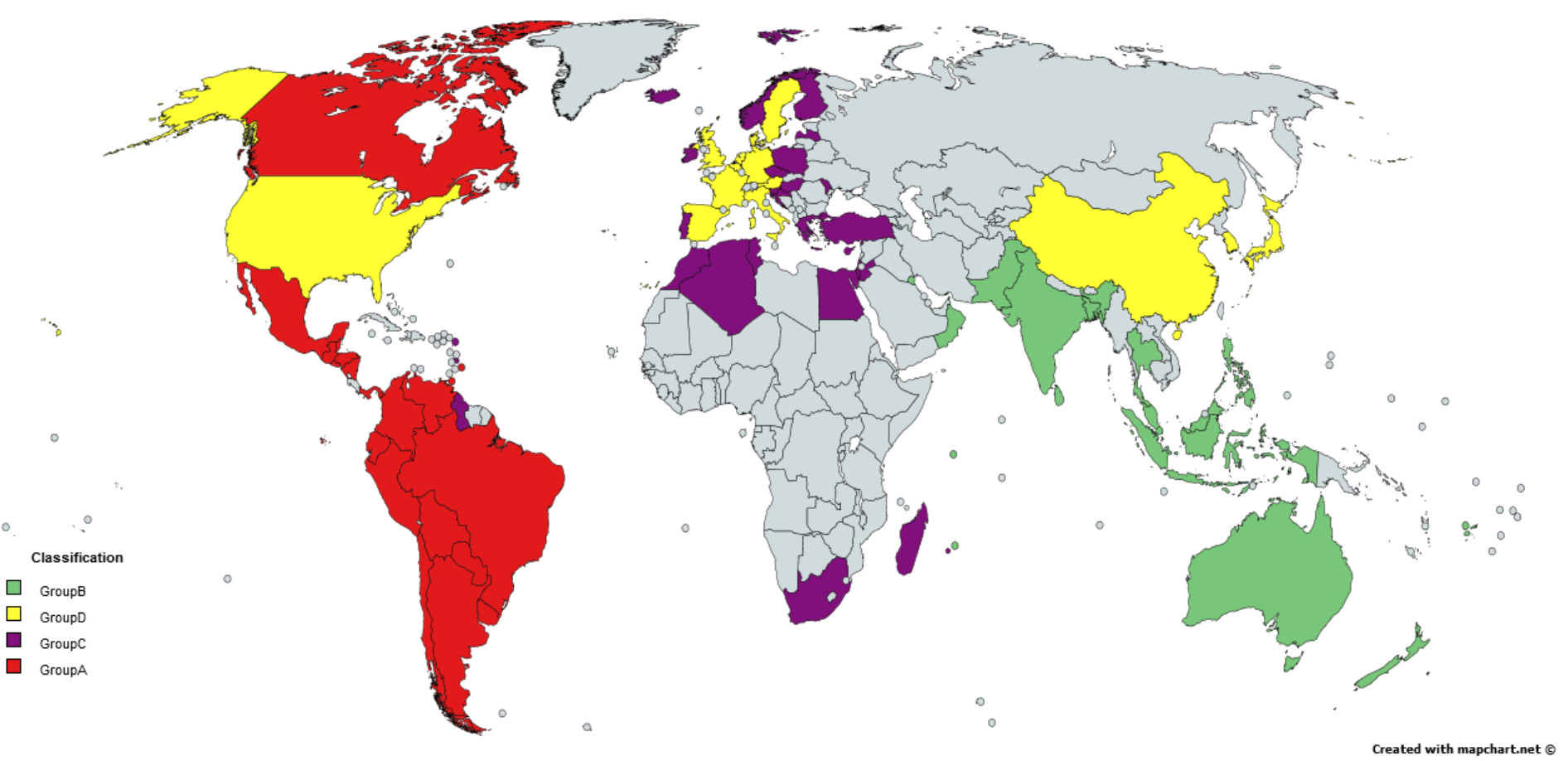}
    \caption{Clustering countries for $k=4$ (New similarity method), made using \url{https://mapchart.net/detworld.html}}
    \label{fig:countries_colorNew}
\end{figure}

Using the new similarity measure, we see that figure \ref{fig:countries_colorNew} is very similar to the figure \ref{fig:countries_color}. The only differences are Argentina and Denmark. Both classifications thus seem almost equivalent.

Our method is more efficient on such graphs because the number of exchanges between members of a same cluster is higher than in the Florida web.

\FloatBarrier
\section{Conclusion}
Let's summarise our main achievements:
\begin{itemize}
    \item The clustering algorithm has a reduced complexity by using $k$-means algorithm.
    \item Comparing our two similarity measures, we concluded that one should choose Browet's method when robustness is required but our method concerning speed.
    \item We analysed the effect of different parameters ($k$, $p_{in}$, $p_{out}$, the number of elements in each cluster and the total number of elements) on the performance of our algorithm. Our best community detection algorithm when $k$ is known is the $k$-means algorithm using additional properties of matrix $X$. 
    \item When $k$ is unknown, the best methods for computation time and accuracy, seems to be the SVD method.
\end{itemize}

Using real networks, we saw some flaws of our algorithm. If the classification of the different countries seems to be accurate, this is not the case for the Florida web. Indeed, the level of perturbation is high compared to the number of edges between clusters ($p_{in} \simeq p_{out}$). For such graphs, we know in advance the reduced graph. We could use this information by checking the structure of the graph after $k$-means by correcting the classification of doing $k$-means again until the classification seems to fit the structure. However, the method would then be very dependant on the type of graph.

\newpage
\section{Appendix}
\subsection{Values of inner product of row vectors of factor matrix $X$ according to different graph}
One can see from the examples below that the distribution of angles of two vectors from different clusters can be highly influenced by the level of noise and form of reduced graph.
\begin{equation*}
   B=
 \begin{pmatrix}
0 & 1 & 0 & 0 & 0\\
0 & 0 & 1 & 0 & 0\\
1 & 0 & 0 & 0 & 0\\
0 & 0 & 0 & 1 & 0\\
0 & 0 & 0 & 0 & 1\\
\end{pmatrix}
\end{equation*}
\begin{center}
number of elements in every clusters ${10,10,10,10,160}$\\
\end{center}
\includegraphics[scale=0.25]{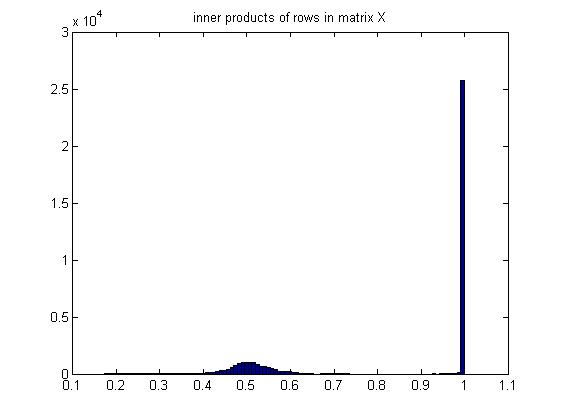}
\includegraphics[scale=0.25]{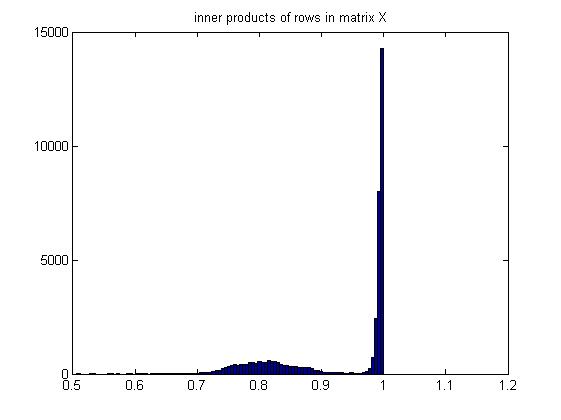}
\includegraphics[scale=0.25]{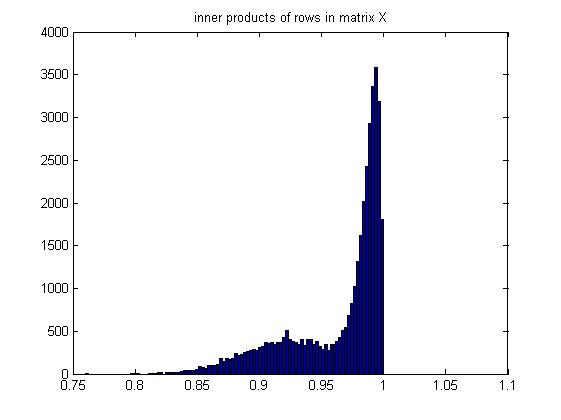}
\begin{equation*}
   B=
 \begin{pmatrix}
1 & 1 & 1 & 0 & 0\\
0 & 0 & 1 & 1 & 0\\
1 & 0 & 0 & 1 & 1\\
0 & 0 & 0 & 1 & 0\\
0 & 0 & 0 & 0 & 1\\
\end{pmatrix}
\end{equation*}
\begin{center}
number of elements in every clusters ${40,40,40,40,40}$
\end{center}
\includegraphics[scale=0.25]{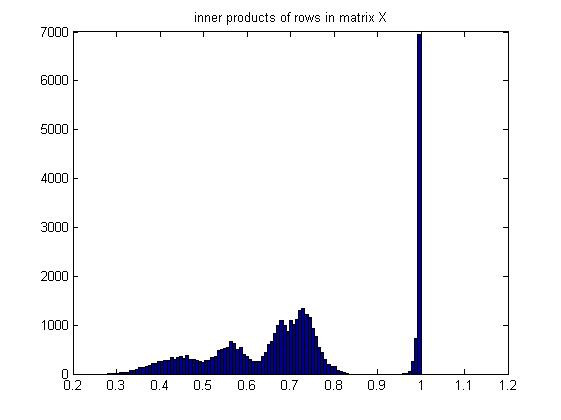}
\includegraphics[scale=0.25]{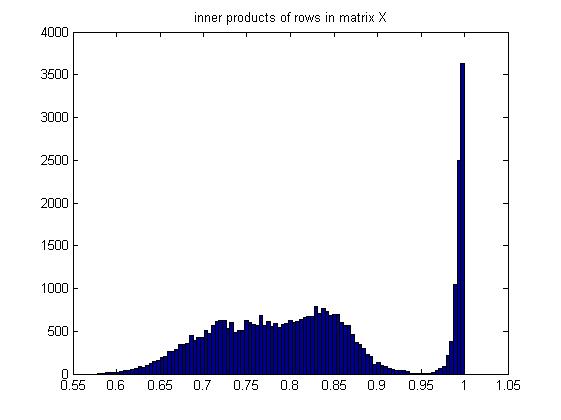}
\includegraphics[scale=0.25]{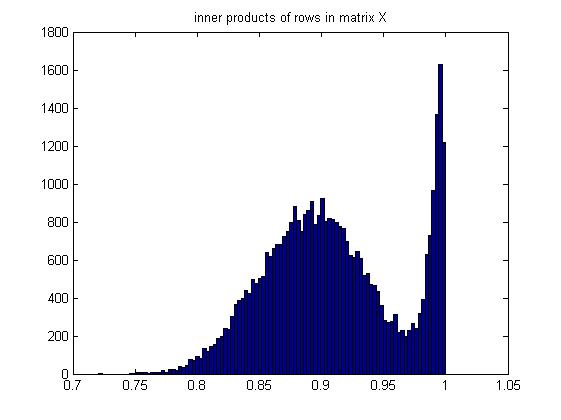}\\
\begin{equation*}
   B=
 \begin{pmatrix}
1 & 1 & 1 & 0 & 0\\
0 & 0 & 1 & 1 & 0\\
1 & 0 & 0 & 1 & 1\\
0 & 0 & 0 & 1 & 0\\
0 & 0 & 0 & 0 & 1\\
\end{pmatrix}
\end{equation*}
\begin{center}
number of elements in every clusters ${10,10,10,10,160}$\\
\end{center}

From the examples above, one can see that when the level of noise is high, there will be less difference between inner product values. Once, the distribution of inner products become totally dense, it will be very difficult to have a good classification. We have to mention that for a complex graph structure, even with no noise, the inner product of two vectors from different clusters can be nonzero.
 
\FloatBarrier


\newpage

\begin{thebibliography}{9}
\bibitem{IEEEhowto:Browet1}
Browet, A., \& Van Dooren, P. (2013). Low-rank Similarity Measure for Role Model Extraction. arXiv preprint arXiv:1312.4860.", \url{http://arxiv.org/abs/1312.4860}

\bibitem{IEEhowto:BrowetThesis}
Browet, A. (2014). \textit{ Algorithms for community and role detection in networks} (Doctoral dissertation, UCL)., \url{http://perso.uclouvain.be/arnaud.browet/files/thesis/thesis.pdf}

\bibitem{IEEhowto:Kmeans1}
Telgarsky, M., \& Vattani, A. (2010). Hartigan's Method: k-means Clustering without Voronoi. In \textit{ AISTATS} (pp. 820-827).,
\url{http://jmlr.csail.mit.edu/proceedings/papers/v9/telgarsky10a/telgarsky10a.pdf}

\bibitem{IEEhowto:KmeansCompl}
Andrea L., (2008), Clustering algorithms : K-means,  [Powerpoint slides], Princeton, Retrieved from  \url{http://www.cs.princeton.edu/courses/archive/spr08/cos435/Class_notes/clustering2_toPost.pdf}

\bibitem{IEEhowto:Kmeans++}
Arthur, D., \& Vassilvitskii, S. (2007, January). \textit{ k-means++: The advantages of careful seeding}. In Proceedings of the eighteenth annual ACM-SIAM symposium on Discrete algorithms (pp. 1027-1035). Society for Industrial and Applied Mathematics., 
\url{http://ilpubs.stanford.edu:8090/778/1/2006-13.pdf}

\bibitem{IEEhowto:NMI}
Strehl, A., \& Ghosh, J. (2002).  \textit{Cluster ensembles---a knowledge reuse framework for combining multiple partitions.} Journal of machine learning research, 3(Dec), 583-617.
\url{http://www.jmlr.org/papers/v3/strehl02a.html}
\end{thebibliography}
\end{document}